\documentclass[aps, pre, showpacs, twocolumn,floatfix, amsmath,amssymb,groupedaddress,letterpaper]{revtex4-1}
\usepackage{graphicx} 
\usepackage{color}
\usepackage{subfigure}

\begin{document}

\title{Neuron dynamics in the presence of $1/f$ noise}
\author{Cameron Sobie}
\email[]{csobie@uvic.ca}
\author{Arif Babul}
\email[]{babul@uvic.ca}
\author{Rog\'{e}rio de Sousa}
\email[]{rdesousa@uvic.ca}
\affiliation{Department of Physics and Astronomy, University of Victoria,
Victoria, B.C., V8W 3P6, Canada}

\date{\today}

\begin{abstract}
  Interest in understanding the interplay between noise and the
  response of a non-linear device cuts across disciplinary boundaries.
  It is as relevant for unmasking the dynamics of neurons in noisy
  environments as it is for designing reliable nanoscale logic circuit
  elements and sensors.  Most studies of noise in non-linear devices
  are limited to either time-correlated noise with a Lorentzian
  spectrum (of which the white noise is a limiting case) or just white
  noise.  We use analytical theory and numerical simulations to study
  the impact of the more ubiquitous ``natural'' noise with a $1/f$
  frequency spectrum.  Specifically, we study the impact of the $1/f$
  noise on a leaky integrate and fire model of a neuron.  The impact
  of noise is considered on two quantities of interest to neuron
  function: The spike count Fano factor and the speed of neuron
  response to a small step-like stimulus. For the perfect (non-leaky)
  integrate and fire model, we show that the Fano factor can be
  expressed as an integral over noise spectrum weighted by a (low
  pass) filter function given by ${\cal F}(t,f)={\rm sinc}^{2}(\pi f
  t)$.  This result elucidates the connection between low frequency
  noise and disorder in neuron dynamics.  Under $1/f$ noise, spike
  dynamics lacks a characteristic correlation time, inducing the leaky
  and non-leaky models to exhibit non-ergodic behavior and Fano factor
  increasing logarithmically as a function of time. We compare our
  results to experimental data of single neurons in vivo [M.C. Teich,
  C. Heneghan, S.B. Lowen, T. Ozaki, and E. Kaplan, Journal of the
  Optical Society of America A {\bf 14}, 529 (1997)], and show how the
  $1/f$ noise model provides much better agreement than the usual
  approximations based on Lorentzian noise. The low frequency noise,
  however, complicates the case for information coding scheme based on
  interspike intervals by introducing variability in the neuron
  response time. On a positive note, the neuron response time to a
  step stimulus is, remarkably, nearly optimal in the presence of
  $1/f$ noise. An explanation of this effect elucidates how the brain
  can take advantage of noise to prime a subset of the neurons to
  respond almost instantly to sudden stimuli.
\end{abstract}


\maketitle

\section{Introduction}

One of the major puzzles of neuroscience is how neurons can store,
process, and compute despite the fact that the brain is extremely
noisy \cite{koch_book}.  Understanding the evolved mechanisms and the
associated non-linear dynamics that allow the neurons to function in
-- and even exploit -- a noisy environment is an essential step
towards gaining insight into the information transmission and
communication networks in the brain.  Such studies also have important
implications beyond the domain of biophysics and neuroscience.  Noise
provides a critical barrier for the development of sensitive
electronic and mechanical devices, particularly at the nanoscale
\cite{weissman88,kogan96}. Increasingly, researchers are focusing on
exploring innovative, non-traditional device design and control
strategies that exploit the ambient noise
\cite{badzey05,almog07,murali09,guerra09,guerra10,zamora10,worschech10,fierens10}.
In this regard, there are clear advantages to understanding how nature
has managed to harness noise in a setting whose primary (apparent)
function is to manage information.

It is generally accepted that neurons communicate with each other
using sharp electric pulses referred to as action potentials or
spikes. Each neuron is connected to several other neurons, and will
only generate a spike output when the integrated input from other
neurons exceeds a certain threshold
\cite{gerstein64,hodgkin52,lapicque07}.  A startling discovery that
(under certain circumstances) neurons can spike more regularly when
stimulated by noise \cite{longtin91,douglass93,mainen95} led to
assertions that noise is inherent to neuron function.  Several
subsequent experimental and theoretical studies were aimed at
elucidating the functionality of neural noise
\cite{nozaki99b,nozaki99a,soma03,middleton03,rossum03,yu05}.  In the
first instance, noise -- as expected -- introduces a variability in
the interspike intervals and degrades the information capacity of the
spike trains, with low contrast signals being most affected
\cite{rossum03}. At the same time, these studies also found that
stochastic resonance provides a mechanism for neurons to take
advantage of their own noise.  In stochastic resonance, the addition
of an appropriate amount of noise in a non-linear system can induce
regularity by sensitizing subthreshold excitations, thus providing the
extra energy for them to reach threshold
\cite{longtin91,douglass93,nozaki99a} and enabling their detection.
Additionally, Brunel {\it et al.} \cite{brunel01} and Svirskis
\cite{svirskis03} have shown that a model neuron, when subjected to
low frequency noise, is able to respond faster to a sudden excitation
than in the absence of noise.  For an animal living in a natural
environment, the ability to react quickly to sudden threats can mean
the difference between life and death.

All of the studies to date that have considered the impact of low
frequency noise on neurons tend to model noise characterized by a
single Lorentzian power spectrum. Natural noise, however, has an
ubiquitous $1/f$ frequency dependence
\cite{kogan96,press78,voss75,hateren97,hausdorff96,diba04}.
From current carrying electronic devices and geophysical time series
to biological systems, the $1/f$ power spectrum is everywhere. In
biological settings, human hearing and speech \cite{voss75}, the response of
biological photoreceptors to large intensity variation of visual image
streams in nature \cite{hateren97}, the stride intervals time series of normal
human gait \cite{hausdorff96}, intrinsic noise in neuronal membranes due to
stochastic opening and closing of the various ion channels \cite{diba04}, etc.
all exhibit $1/f$ behavior.  Guerra {\it et al.} \cite{guerra09} demonstrated how
stochastic resonance induced by $1/f$ noise can increase the sensitivity
of nanomechanical resonators, allowing for the possibility of
fashioning them into noisy but robust nanoscale computation devices.
Neither the response nor the details of the underlying non-equilibrium
behavior that makes neurons robust to natural ($1/f$) noise is well
understood.


In this paper, we present a comprehensive study of how neuron dynamics 
is affected by an
arbitrary noise spectral density, and which sectors of the spectra are
responsible for the beneficial functions that noise can provide.
Specifically, we explore whether neuron response under
$1/f$ noise is significantly different from the one found in the
presence of a simple Lorentzian spectra.  

The relevance of low frequency noise, implied by a $1/f$ spectrum, to
spike dynamics at the single neuron level is especially evident, as we
shall argue, in the direct \emph{experimental} measurements of the
spike count Fano factor by Teich {\it et al.}\cite{teich97,turcott95}
(Fano factor is the ratio between the variance and the mean spike
number during a given observation time).  These authors demonstrated
that the Fano factor of single neurons in the visual systems of cats
and insects increases monotonically as a function of time.  This is in
dramatic contrast to the simple Poisson model (white noise), which
leads to Fano factor equal to one at all times.  The monotonic rise is
also incompatible with models based on Lorentzian noise because the
resulting Fano factor saturates at times longer than the inverse
Lorentzian half width \cite{middleton03,schwalger08}.  We show that
the characteristic non-ergodicity of $1/f$ noise explains why the Fano
factor never saturates in single neuron experiments. Moreover, the
rate at which the Fano factor grows as a function of time is different
for $1/f$ and Lorentzian.

In addition, we consider the effect of $1/f$ noise on the reaction
time of a neuron in response to a sudden stimulus. We demonstrate that
$1/f$ noise is \emph{nearly optimal} for speeding up neuron response. We
provide an explanation for this effect that sheds light on the
mechanism of neuron adaptation to their noisy environment.

This paper is organized as follows. Section~\ref{lif} describes our
model for the neuron: the leaky integrate and fire (LIF) model of the
neuron, and explains how we introduce $1/f$ noise and other spectral
densities in this model.  Section~\ref{fanosec} describes an
analytical theory and a set of numerical simulations of the neuron
Fano factor as a function of time and compares our result to
experiment \cite{teich97}. Particularly notable is our general
expression Eq.~(\ref{fnoise}) relating the Fano factor to an integral
over low frequency noise weighted by an appropriate filter function.
Section~\ref{suddensec} addresses the question of how noise can
provide a mechanism for neurons to respond faster to a sudden
stimulus. Section~\ref{concsec} provides our concluding remarks.

\section{The LIF model \label{lif}}

From a biophysical perspective, the classical Hodgkin-Huxley model
\cite{hodgkin52} and its contemporary variants represent the most realistic
mathematical description of electrical response of a single neuron.
Due to their intrinsic complexity, however, such models render the
theoretical and computational analysis of neuronal and neural network
dynamics exceedingly difficult.  For this reason, most studies to date
tend to reference the simpler spiking neuron models, of which the
leaky integrate and fire (LIF) model \cite{gerstein64,lapicque07} that
we adopt is one.  The LIF model represents each neuron by an
electrical circuit; when appropriate circuit parameters and features
are chosen, the LIF model can reproduce quite similar dynamics to the
one described by the more complex Hodgkin-Huxley model
\cite{koch_book}.

The LIF model consists of a capacitor $C$ in parallel with a resistor
$R$; an injected continuous current $I(t)$ models the spike input from
a large number of neighboring neurons.  The neuron (or capacitor)
voltage $V(t)$ is given by the circuit equation,
\begin{equation}
C\frac{dV(t)}{dt}+\frac{V(t)}{R}=I(t).
\label{dvdt}
\end{equation}
A spike is generated whenever the voltage across the capacitor reaches
a certain threshold $V_{\rm{th}}$; after the spike is emitted, the
neuron is reset to a zero voltage state.  Note how the threshold rule
for spike generation introduces non-linearity in the LIF model:
Consider two input currents $I_1(t)$ and $I_2(t)$; a neuron subject to
input $[I_1(t) + I_2(t)]$ will generally reach threshold faster than a
neuron that is subject to either $I_1(t)$ or $I_2(t)$ only. Hence,
the sum of outputs obtained from $I_1(t)$ and $I_2(t)$ applied separately
is different from the output obtained from $[I_1(t) + I_2(t)]$.
Also, the resistance $R$ plays an important role in the model: it
allows charge to leak out, thus negating inputs received in the
distant past.  Neurons have the property that inputs long past have
less effect than recent inputs; a sufficient amount of input must
happen sufficiently rapidly for the neuron to fire.

The version of the LIF model that we are using has an additional
feature that makes it more realistic: The introduction of a refractory
time period $\tau_r$, which models the physical reset time for a
neuron after emitting a spike.  This prevents the neuron from
receiving input for a time $\tau_r$ after spiking. Our choice for
these circuit parameters are given in Table~\ref{table1}.

The presence of a leak and a refractory period makes the LIF neuron
extremely hard to treat analytically \cite{schwalger08}.  As a result,
many theoretical studies have focused on the $R=\infty$ and $\tau_r=0$
limit, the so called perfect (non-leaky) integrate and fire model
\cite{middleton03, lindner04}.  This latter model is much easier to analyze but as 
we will demonstrate, the reduced complexity also leads to significantly 
different dynamics.

\subsection{Introducing noise in the LIF model\label{intronoise}}

We considered the LIF model subject to a noisy input current of the form
\begin{equation}
I(t)=\theta\left(I_0 + I_1 \eta(t)\right),
\label{currit}
\end{equation}
where $I_0$ is a (constant) bias current, $I_1$ is the noise
amplitude, and $\theta(x)$ is the Heaviside step function:
$\theta(x)=1$ for $x\geq 0$ and $\theta(x)=0$ for $x< 0$; this ensures
the current input represents the sum of spikes from a large number of
connected neurons.  The time series $\eta(t)$ is a Gaussian stochastic
process, with variance equal to one and power spectra given by
\begin{equation}
\tilde{S}(f)=\frac{1}{2\pi}\int_{-\infty}^{\infty} dt \;\textrm{e}^{i 2\pi f t} \langle \eta(t)\eta(0)\rangle,
\label{noisespec}
\end{equation}
with the brackets $\langle\quad...\quad\rangle$ denoting ensemble averages over a
large number of time series $\eta(t)$. Appendix A describes the method
used to generate individual time series for any given noise spectral
density $\tilde{S}(f)$.  We considered a number of different noise
densities, including the family of power law spectral densities
\begin{equation}
\tilde{S}_{\alpha}(f)=A_{\alpha} \frac{1}{f^{\alpha}},
\label{aof}
\end{equation}
where $\alpha$ is an exponent ($\alpha=1$ corresponds to $1/f$ noise).
The normalization constant $A_{\alpha}$ is set by the condition for
the variance to be one, $2\pi\int df \tilde{S}(f)=\langle
\eta^{2}\rangle=1$ [In the case of $\alpha=1$ Eq.~(\ref{aof}) is valid
for $\gamma_{\rm{min}}<f<\gamma_{\rm{max}}$; $\gamma_{\rm{min}}$ is a
lower cut-off for which $\tilde{S}_{\alpha}(f)$ saturates, and
$\gamma_{\rm{max}}$ is an upper cut-off for which
$\tilde{S}_{\alpha}(f)$ goes to zero faster than $1/f^2$ -- see below].

Another important class of noise spectral density arises when the environment fluctuates with a single 
characteristic time $\tau_c$:
\begin{equation}
\tilde{S}(f)= \frac{1}{2\pi^2} \frac{\gamma}{f^{2}+\gamma^{2}},
\label{lornoise}
\end{equation}
with $\gamma\equiv 1/(2\pi\tau_c)$.  This is a Lorentzian power spectrum and it implies $S(t)=\langle
\eta(t)\eta(0)\rangle=\textrm{e}^{-|t|/\tau_c}$.  Many authors
refer to $\tau_c$ as the ``correlation time'', and to the Lorentzian
spectra as ``time correlated noise''.  In the
limit that $\gamma$ goes to infinite ($\tau_c\rightarrow 0$),
$\tilde{S}(f)\approx (2\pi^2 \gamma)^{-1}$ is approximately constant
for all $f\ll\gamma$.  Hence $\gamma\rightarrow \infty$ is the ``white
noise'' limit.  Another important limit occurs when $\gamma\rightarrow
0$ ($\tau_c\rightarrow \infty$): In this case $\tilde{S}(f)\rightarrow
\frac{1}{2\pi}\delta(f)$, signaling a ``static'' limit.

It is useful to recall the basic physical picture for the origin of
$1/f$ noise.  It emerges from the combination of a large number of
Lorentzian fluctuators with an exponentially wide distribution of
characteristic rates $\gamma$ \cite{kogan96,weissman88}. For example, assume
$\gamma = \gamma_{\rm{max}} \textrm{e}^{-\lambda }$, with $\lambda$ a random
variable that represents a distribution of activation energies.
Assuming $\lambda$ is uniformly distributed in the interval
$[0,\lambda_{\rm{max}}]$ we get
\begin{eqnarray}
\tilde{S}(f)&=&\int_{0}^{\lambda_{\rm{max}}}\frac{d\lambda}{\lambda_{\rm{max}}}\frac{1}{2\pi^2}\frac{\gamma}{f^2+\gamma^2} \nonumber\\
&&=\frac{1}{2\pi^2\lambda_{\rm{max}}}\int_{\gamma_{\rm{min}}}^{\gamma_{\rm{max}}}\frac{d\gamma}{\left|\frac{d\gamma}{d\lambda}\right|}\frac{\gamma}{f^2+\gamma^2}\label{oofnoise1}\\
&=&\frac{\arctan\left(\frac{\gamma_{\rm{max}}}{f}\right)-\arctan\left(\frac{\gamma_{\rm{min}}}{f}\right)}{2\pi^2\lambda_{\rm{max}}}
\frac{1}{|f|}.\nonumber
\end{eqnarray}
When $\gamma_{\rm{min}}\ll |f|\ll \gamma_{\rm{max}}$, we may approximate $\arctan(\gamma_{\rm{max}}/f)\approx \pi/2$ and 
$\arctan(\gamma_{\rm{min}}/f)\approx 0$; this leads to
\begin{equation}
\tilde{S}(f)\approx \frac{1}{4\pi \ln\left(\frac{\gamma_{\rm{max}}}{\gamma_{\rm{min}}}\right)}\frac{1}{|f|},
\label{oofnoise}
\end{equation}
where we used the fact that 
$\lambda_{\rm{max}}=\ln(\gamma_{\rm{max}}/\gamma_{\rm{min}})$.  Hence, overall the resultant noise 
is well described by
\begin{equation}
\tilde{S}(f)=\left\{
\begin{array}{c c}
\frac{A_1}{\gamma_{\rm{min}}}   & 0\leq |f|<\gamma_{\rm{min}}\\
\frac{A_1}{|f|}  &
\gamma_{\rm{min}} \leq |f| < \gamma_{\rm{max}}\\
0 & \gamma_{\rm{max}} \leq |f|<\infty
\end{array}
\right.,
\label{oneof}
\end{equation}
with constant $A_1=\left[4\pi \ln\left(\gamma_{\rm{max}}/\gamma_{\rm{min}}\right)\right]^{-1}$.

From Eq.~(\ref{oofnoise1}) we see that the distribution of Lorentzian
linewidths $\gamma$ is given by
$P(\gamma)=1/|d\gamma/d\lambda|=1/\gamma$. This is the reason why the
spectrum acquires the $1/f$ dependence. We may generalize this
distribution to $P(\gamma)=1/\gamma^{\alpha}$, with $\alpha$ a
dimensionless exponent; carrying though a similar derivation as in
Eq.~(\ref{oofnoise1}) leads to $\tilde{S}(f)\propto 1/f^{\alpha}$. This
shows that deviations of the $1/\gamma$ distribution will reflect
directly into a $\alpha\neq 1$ exponent for the noise spectrum.

Usually, $\gamma_{\rm{min}}$ is exponentially small, and the
experimental observation time window $T$ is \emph{smaller} than
$\gamma_{\rm{min}}^{-1}$. In this case the low frequency cut-off will
be instead set by $\gamma_{\rm{min}}=T^{-1}$.  As the observation time $T$ increases,
more low frequency fluctuators will play a role; as a result, $1/f$
noise has no characteristic time scale, and displays
\emph{non-ergodic} behavior (time averages of observables are
non-convergent, and can not be equivalent to ensemble averages).
Below we discuss how the non-ergodic property leads to an increasing
neuron Fano factor as a function of time.

\subsection{White versus $1/f$ noise in the superthreshold regime: Bursting phenomena}

\begin{figure*}
  \subfigure[White Noise]{\label{fig:wtime1}\includegraphics[width=0.49\textwidth]{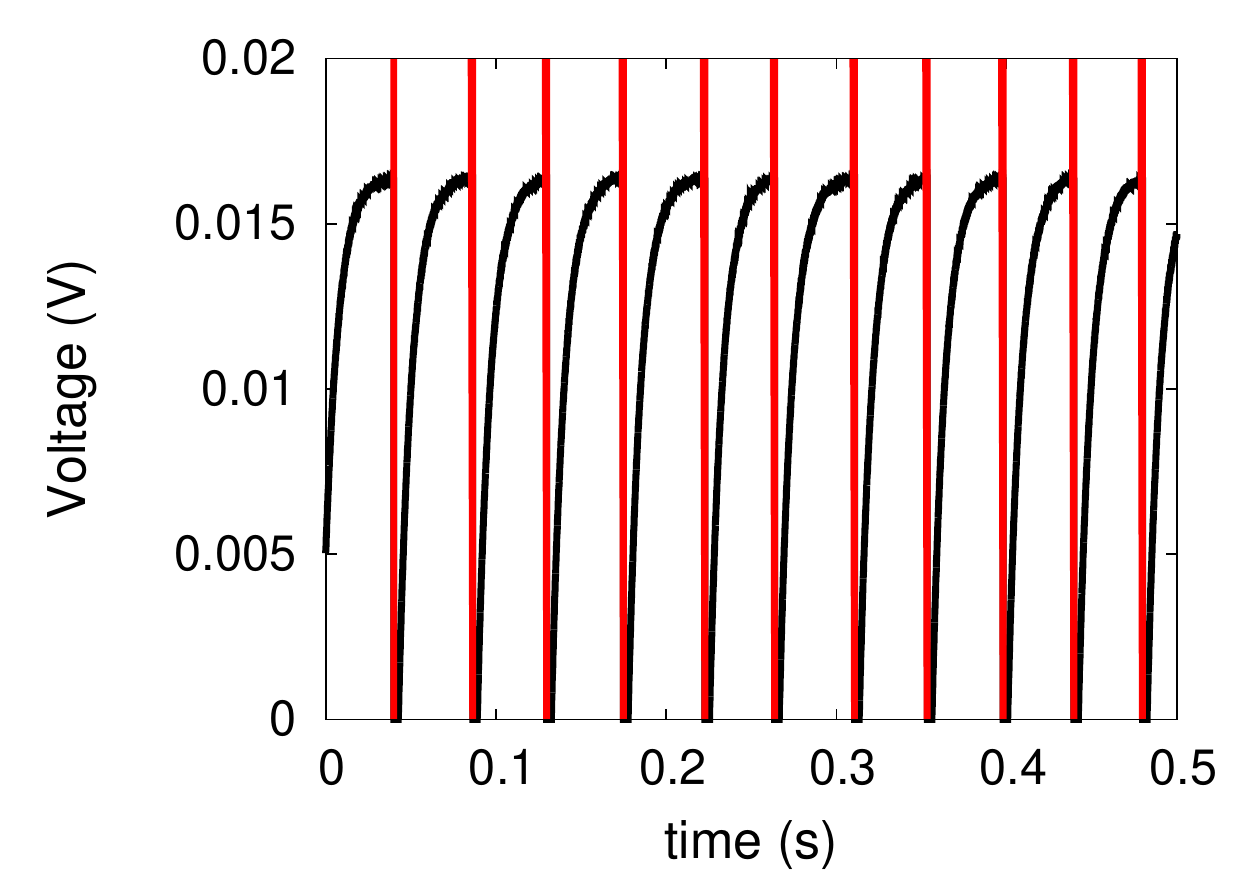}}
  \subfigure[$1/f$ Noise]{\label{fig:ptime1}\includegraphics[width=0.49\textwidth]{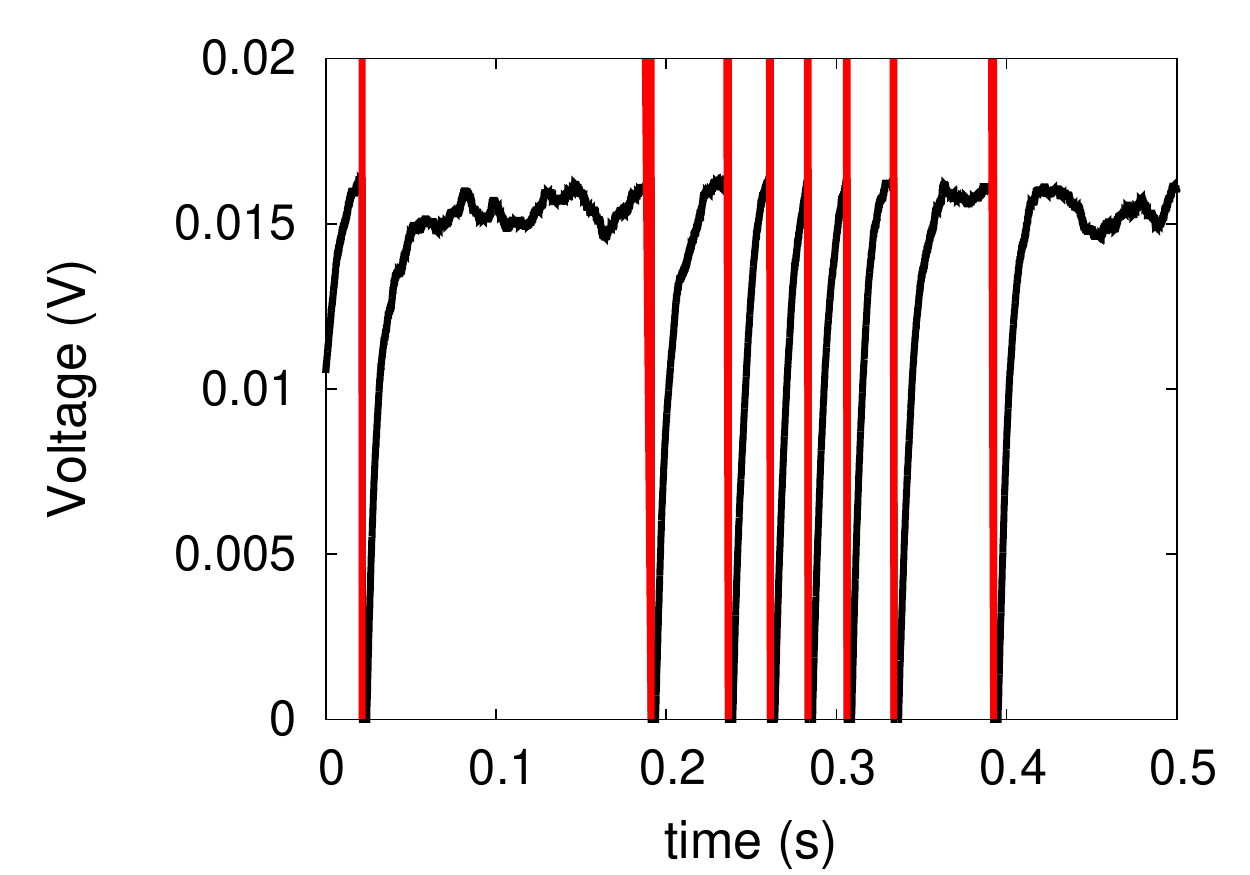}}
  \caption{(color online). Neuron voltage as a function of time (black
    curve), for (a) white noise and (b) $1/f$ noise.  The vertical
    lines (red) denote spiking events. Note how $1/f$ noise leads to
    long time intervals with no spike generation, followed by
    intervals with spike bursting.\label{fig:timeseries}}
\end{figure*}

To compare LIF dynamics under the effect of white and $1/f$ spectra,
we considered noise in the superthreshold regime
($I_0>V_{\rm{th}}/R=4.28\times 10^{-10}$~A). We assumed
$I_0=4.3\times 10^{-10}$~A, slightly above threshold, ensuring that
without noise the neuron will spike every $46$~ms. For the cases with
noise, we used $I_1=0.1 I_0$.  The algorithm of Appendix A was used to generate a
current input $I(t)$, and Eq.~(\ref{dvdt}) is integrated using the
Runge-Kutta method. Figures~\ref{fig:wtime1}~and~\ref{fig:ptime1}
shows the neuron voltage as a function of time, for a particular time
series (the observation time window was $T=2$~s).  Under white noise, the spiking remains quite regular over
time, because $I(t)$ varies rapidly and most of its fluctuating
components are filtered out. On the other hand, $1/f$ noise shows a
combination of long periods of inactivity, with the voltage taking a
long time to reach threshold, together with periods of spike bursting
where the voltage reaches threshold on a much shorter time scale.
This is a result of the fact that under $1/f$ noise, the current tends
to get ``stuck'' at either small or large values.

\begin{table}
  \caption{Circuit parameters used in our LIF model. The parameters are similar to the 
    ones used to describe neurons in the cat's visual cortex (Chapter 14 of \cite{koch_book}).\label{table1}}
\begin{center}
\begin{tabular}{|c|l|}
\hline
\hline
Resistance & $R=38.3$~M$\Omega$\\
\hline
Capacitance & $C=0.207$~nF\\
\hline
Circuit time constant & $RC = 7.93$~ms\\
\hline
Threshold voltage & $V_{\rm{th}}=16.4$~mV\\
\hline
Refractory period & $\tau_r=2.68$~ms\\
\hline
\hline
\end{tabular}
\end{center}
\end{table}

\begin{figure}
\includegraphics[width = 0.49\textwidth]{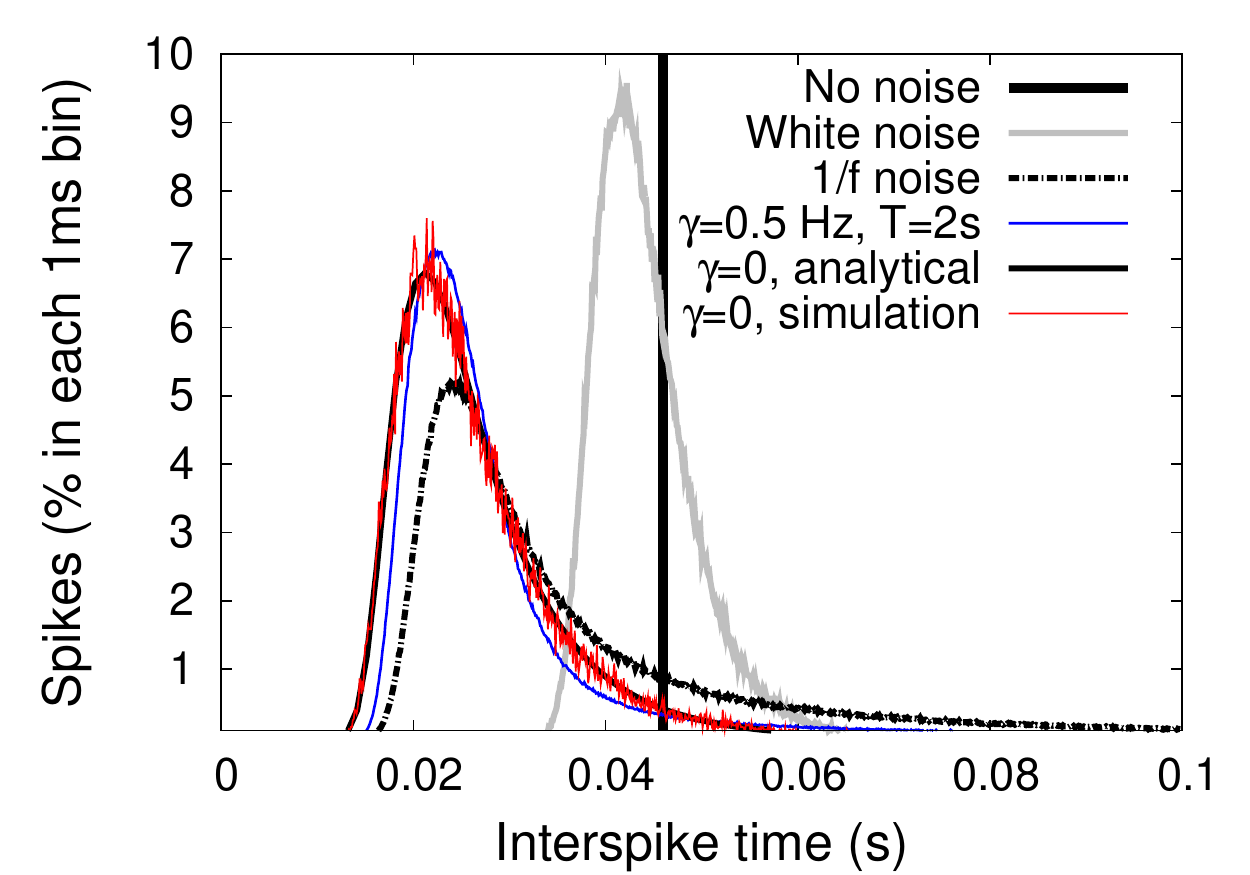}
\caption{(color online) Interspike interval histogram (ISI) for the cases of white
  and $1/f$ noise considered in
  Figs.~\ref{fig:wtime1}~and~\ref{fig:ptime1}, respectively.  The
  noise amplitude was $I_1=0.1 I_0$, with bias current $I_0$ slightly above
  threshold. Also shown is the case of no noise ($I_1=0$), and of Lorentzian noise 
  with $\gamma=0$~Hz and $\gamma=1/T=0.5$Hz, where $T=2$~s was 
  the simulation time window for each time series.
  Note how $1/f$ has much broader ISI, with a long time tail extending well
  beyond its mean interspike time.\label{fig:ISI}}
\end{figure}

Figure~\ref{fig:ISI} shows the interspike time interval histogram
(ISI) for a 100,000 ensemble of time series with the same parameters
considered in Figs.~\ref{fig:wtime1}~and~\ref{fig:ptime1}.  Here we
see that both types of noise can cause a notable decrease in the mean
interspike time interval (compare to the noiseless case of a constant
bias current). However, 1/$f$ noise leads to a far more dramatic
shift. A large portion of the interspike time histogram lies
significantly below the noiseless interval, and a long tail is
observed at large interspike times.  This behavior is characteristic
of bursting (See e.g. Chapter 16 of \cite{koch_book}); several spikes
occur in rapid succession, followed by a longer period without spike
activity.

Figure ~\ref{fig:ISI} also shows the ISI for two Lorentzian noise
spectra, with $\gamma=0$~Hz and $\gamma=1/T=0.5$~Hz, where $T=2$~s was
the simulation time window for each time series.  Lorentzian noise
with such low frequency leads to an ISI that is nearly as broad as the
$1/f$ noise case; however, the Lorentzian noise cases do not display
the long time tail characteristic of $1/f$ noise. The simulations for
$\gamma=0$~Hz can be compared to an exact analytical result obtained
using the methods of Refs.~\cite{middleton03,lindner04} (See
Appendix~\ref{appendixb}). Note how the $\gamma=0$~Hz simulations are
in excellent agreement with the exact result.

\section{Neuron Fano factor under $1/f$ noise \label{fanosec}}

\subsection{Fano factor in the perfect integrate and fire model: Analytical results\label{fanoanal}}

The Fano factor is defined by\cite{koch_book}
\begin{equation}
F(t)=\frac{\langle \left[\Delta N(t)\right]^{2}\rangle}{\langle N(t)\rangle},
\label{fano}
\end{equation}
where the random variable $N(t)$ is the number of spikes generated
from $t'=0$ to $t'=t$, and $\Delta N(t)=N(t)-\langle N(t)\rangle$.
Hence, the Fano factor measures the amount of uncertainty in the spike
train at a given time $t$. A noiseless spike train with identical
interspike time intervals yields $F(t)=0$. In contrast, consider the
case that the spike events are uniformly distributed in the interval
$[0,t]$: In this case the probability for a spike event to happen
during a time interval $[t',t'+\Delta t]$ is independent of $t'$ and
given by $\mu \Delta t$, where $\mu$ is the mean firing rate.  Then
the resulting $N(t)$ is a Poisson random variable uncorrelated in time
(white noise) with $\langle N(t)\rangle=\langle(\Delta
N)^{2}\rangle=\mu t$, leading to $F(t)=1$ at all times.  In the
presence of low frequency noise, $F(t)$ is known to become larger than
one \cite{middleton03,schwalger08}.

The observation of a Fano factor much larger than one rules out the
simple Poisson model, and suggests the presence of long time
correlations in the data \cite{teich97,koch_book}.  Middleton {\it et
  al.} \cite{middleton03} derived an analytic expression for the Fano
factor of the perfect ($R=\infty$) integrate and fire model subject to Lorentzian
noise [Eq~(\ref{lornoise})]. At times much longer than the average interspike interval their result becomes
\begin{equation}
  F_{\rm{Lor.}}(t)=\frac{\langle (\Delta I^2)\rangle}{ \langle I\rangle}\frac{2\tau_c}{CV_{\rm{th}}}
  \left [ 1-\frac{\tau_c}{t}(1-\textrm{e}^{-\frac{t}{\tau_c}}) \right ],
\label{fanoanaly}
\end{equation} 
where $\langle(\Delta I^2)\rangle= I_{1}^{2}$ is the current variance,
$\langle I\rangle= I_0$ is the bias current, $\tau_c$ is the
correlation time of the Lorentzian noise, $C$ is the capacitor's
voltage, and $V _{\rm{th}}$ is the threshold voltage. We emphasize
that Eq.~(\ref{fanoanaly}) assumes an input current $I(t)=I_0 + I_1
\eta(t)$, i.e., it neglects the step function used in our numerical
computations [compare to Eq.~(\ref{currit})].


Here we generalize this result to an arbitrary noise spectral density.
An RC circuit with no leakage ($R=\infty$) will lead to a capacitor voltage $V$ that always increases with
increasing time. In the case of the neuron, $V$ is reset to
zero when it reaches the threshold $V_{\rm{th}}$. In other words, the
voltage is decreased by $V_{\rm{th}}$ each time the neuron
spikes. An equivalent way to treat this reset process is to instead
\emph{increase the threshold} by an additional $V_{\rm{th}}$ each time
the neuron spikes; this allows us to count the number of spikes at a given time $t$ by simply 
dividing the monotonically increasing $V$ by $V_{\rm{th}}$. Therefore
the random variable $N(t)$ is well approximated by
\begin{equation}
N(t)\approx \frac{V(t)}{V_{\rm{th}}}=\frac{1}{CV_{\rm{th}}} \int_{0}^{t} dt' I(t').
\label{ntapprox}
\end{equation}
This approximation is valid at long times, $t\gg CV_{\rm{th}}/I_0$,
i.e.  times much longer than the mean interspike interval so that
$N(t)$ can be represented by a real number instead of an integer.

The simplicity of the perfect integrate and fire model lies in the
fact that we do not need to consider the threshold barrier explicitly.
This property relies heavily on the fact that the voltage of an RC
circuit never decreases when $R=\infty$.  It is worthwhile to show how
the approximation Eq.~(\ref{ntapprox}) fails in the presence of any
amount of leakage. 

When $R<\infty$, the mean voltage according to
Eq.~(\ref{dvdt}) is given by $R I_0 (1-\textrm{e}^{-t/RC})$.
Hence when $t\rightarrow \infty$ the mean voltage saturates at $RI_0$,
and the approximation of Eq.~(\ref{ntapprox}) would give $\langle
N(t\rightarrow \infty)\rangle = RI_0/V_{\rm{th}}<\infty$.  This is
clearly an unphysical result: If the neuron spikes at least once, it
will spike an infinite number of times when $t\rightarrow \infty$, and
$\langle N(t\rightarrow\infty)\rangle$ must be either $0$ or $\infty$.
This unphysical saturation of $\langle N(t)\rangle$ shows why we must
include the threshold barrier explicitly when dealing with the leaky
model; it also shows why it is so difficult to treat the leaky model
analytically.

In the absence of leakage, we have $\langle N(t)\rangle=
t \;I_0/(CV_{\rm{th}})$, i.e. at long times the number of spikes increases indefinitely
with increasing time. The Fano factor can be calculated explicitly
plugging Eqs.~(\ref{currit})~and~(\ref{ntapprox}) into
Eq.~(\ref{fano}),
\begin{eqnarray}
F(t) &=& \frac{CV_{\rm{th}}}{I_0 t}\frac{I_{1}^{2}}{(CV_{\rm{th}})^2} \int_{0}^{t}dt' \int_{0}^{t}dt''
\langle \eta(t')\eta(t'')\rangle \nonumber\\
&=& \frac{2I_{1}^{2}}{CV_{\rm{th}}I_0}\frac{1}{t} \left[
\int_{0}^{t/2}dT \int_{0}^{2T} d\tau S(\tau) \nonumber\right.\\
&&\left.+ \int_{t/2}^{t}dT \int_{0}^{2(t-T)}d\tau S(\tau)\right].
\label{fanost}
\end{eqnarray}
In the last step we changed the variables to $T=(t'+t'')/2$ and
$\tau=(t'-t'')$, and used the symmetry $S(\tau)=S(-\tau)$.
Eq.~(\ref{fanost}) provides an explicit relationship between the Fano
factor and an arbitrary time correlation function. For example, upon
inserting $S(\tau)=\textrm{e}^{-|\tau|/\tau_c}$ we recover
Eq.~(\ref{fanoanaly}) exactly.

A simpler expression can be derived by inserting $S(t)=2\pi \int df
\textrm{e}^{-i2\pi f t} \tilde{S}(f)$ into Eq.~(\ref{fanost}):
\begin{equation}
F(t)=\frac{2\pi I_{1}^{2}}{CV_{\rm{th}}I_0} \;t\int_{-\infty}^{\infty}df \tilde{S}(f){\cal F}(t,f),
\label{fnoise}
\end{equation}
where we defined the filter function by 
\begin{equation}
{\cal F}(t,f)={\rm sinc}^{2}(\pi f t), 
\label{filterfunc}
\end{equation}
where ${\rm sinc}(x)=\sin{(x)}/x$ is the ``unnormalized sinc function''. 
Hence the Fano factor can be expressed as an integral over low
frequency noise; the filter function ${\cal F}(t,f)$ dictates how much
noise is "allowed in" at each given time $t$. Inspecting
Eq.~(\ref{filterfunc}) shows that it acts as a low pass filter with
bandwidth $\approx 1/(\pi t)$. In the limit $t\rightarrow \infty$, we have
$ft\gg 1$ for all frequencies, and ${\cal F}\approx \delta(f)/t$.  This
leads to a useful result
\begin{equation}
F(t\rightarrow \infty)= \frac{2\pi I_{1}^{2}}{CV_{\rm{th}}I_0} \tilde{S}(0).
\label{finf}
\end{equation}
Therefore, the saturation of the Fano factor (or lack thereof) at long
times is directly proportional to the amount of zero frequency noise.

We can use Eq.~(\ref{fanost}) to find the neuron Fano factor subject
to $1/f$ noise.  From Eq.~(\ref{oofnoise1}) we know that the time
correlation function for $1/f$ noise can be written as
\begin{equation}
  S_{1/f}(t)=\frac{1}{\ln{\left(\frac{\gamma_{\rm{max}}}{\gamma_{\rm{min}}}\right)}} \int_{\gamma_{\rm{min}}}^{\gamma_{\rm{max}}}\frac{d\gamma}{\gamma} \textrm{e}^{-\gamma |t|}.
\end{equation}
Hence the Fano factor for $1/f$ noise can be written as a weighted average of Lorentzian Fano factors,
\begin{eqnarray}
F_{1/f}(t)&=&\frac{2I_{1}^{2}}{C V_{\rm{th}}I_0} \frac{1}{\ln{\left(\frac{\gamma_{\rm{max}}}{\gamma_{\rm{min}}}\right)}} 
\int_{\gamma_{\rm{min}}}^{\gamma_{\rm{max}}}\frac{d\gamma}{\gamma^2} 
\left[1-\frac{1}{\gamma t}\left(1-\textrm{e}^{-\gamma t}\right)\right]\nonumber\\
&\approx& \frac{2I_{1}^{2}}{C V_{\rm{th}}I_0} \frac{1}{\ln{\left(\frac{\gamma_{\rm{max}}}{\gamma_{\rm{min}}}\right)}} 
\frac{t}{2} \left[
\frac{(3-2 C_E)}{2}  -\ln{\left(\gamma_{\rm{min}}t \right)}
\right],
\label{fanooneoverf}
\end{eqnarray}
where $C_E=0.5772$ is the Euler-Mascheroni constant.  The latter
approximation is valid for $\frac{1}{\gamma_{\rm{max}}} \ll t\ll
\frac{1}{\gamma_{\rm{min}}}= T$, where $T$ is the experimental time
window.  Hence $F_{1/f}(t)$ increases monotonically with increasing
time, until it reaches a saturation value around $\sim T/\ln{(T)}$.
This saturation value, however, is artifact of the finite length $T$
of the experiment; it diverges as $T\rightarrow \infty$, i.e., when
longer data sets are acquired. This effect is a manifestation of the
non-ergodicity of $1/f$ noise. In stark contrast, for an experimental
window $T>\tau_{c}$, the Fano factor for the simple Lorentzian noise
saturates at $F_{\rm{Lor.}}=2\tau_c I_{1}^{2}/(CV_{\rm{th}}I_0)$, and
remains at this value regardless of whether $T$ is increased further.
Assuming that this behavior carries over to the Leaky model, it offers
one way to determine whether the neuron noise is better described by
$1/f$ or Lorentzian spectrum. The dependence of $F(t)$ on $t$ is
another potential discriminator, as we shall see below.

Figure~\ref{fig:fanopower} illustrates the relevance of the filter
function Eq.~(\ref{filterfunc}) in quantifying the amount of noise
absorbed by neurons at a given time $t$.  Here we plot the Fano factor
$F(t)$ for several different noise spectra, but choose each noise
power (proportional to $I_{1}^{2}$) so that the Fano factor at a
particular time $t=0.1$~s is identical [$F(0.1~{\textrm s})=10$] for
all noise spectra.  Hence at this particular $t=0.1$~s the amount of
disorder on neuron response is the same even though we are describing
neurons subject to very different dynamical environments.
Nevertheless, at times after $t=0.1$~s the Fano factor
differs considerably for different environments.  This is a direct
consequence of the fact that neurons integrate noise over a bandwidth
$\approx 1/(\pi t)$; hence, as $t$ increases, a neuron absorbs noise over an increasingly narrow frequency range. 
Note also how $F(t)$ for $1/f$ noise depends on the total
observation time window $T$, and how $F(t)$ is sensitive to different noise spectra before $t=0.1$~s.

\begin{figure}
\includegraphics[width = 0.49\textwidth]{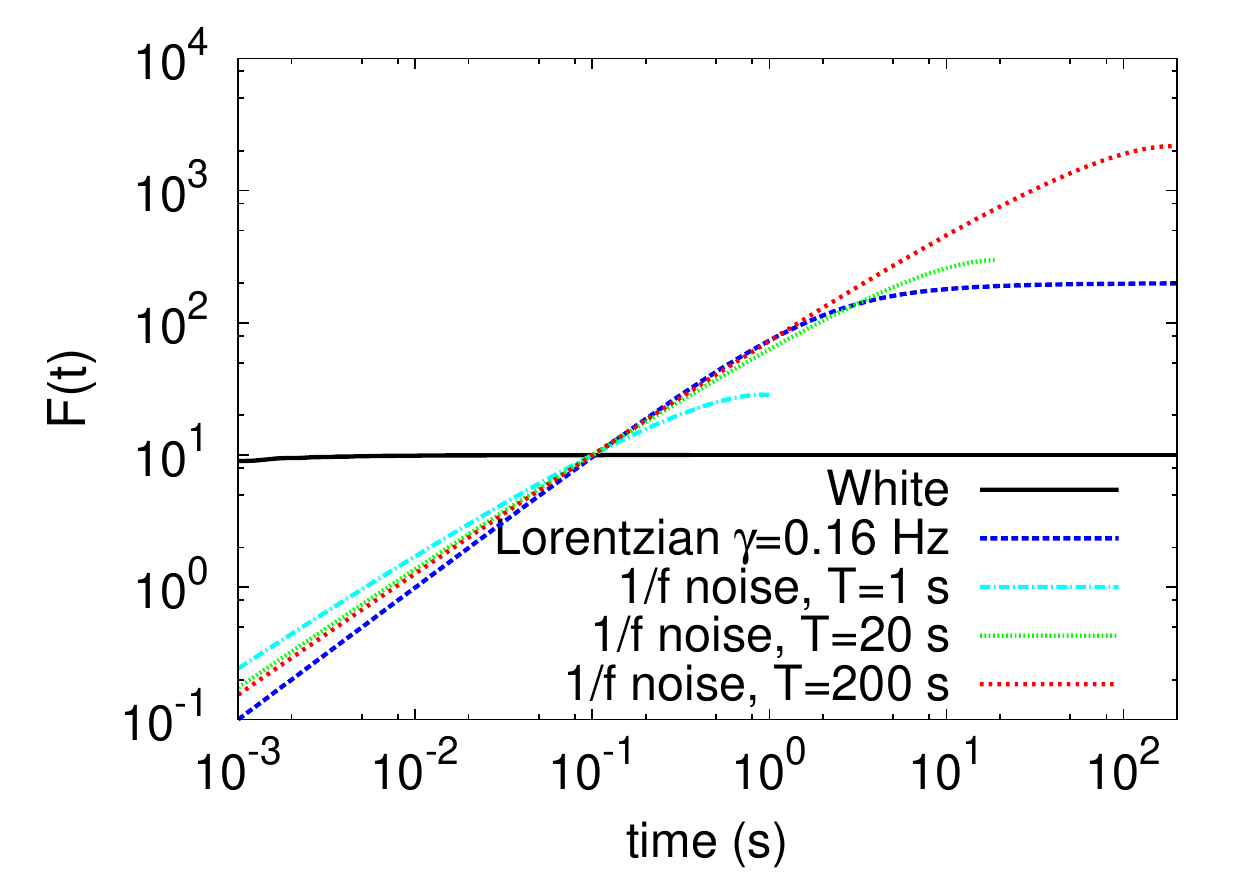}
\caption{(color online) Fano factor [Eq.~(\ref{fano})] for the case
  $R=\infty$ (non-leaky or perfect) integrate and fire model as a
  function of time for white noise, Lorentzian noise with
  $\gamma=0.16$~Hz, and $1/f$ noise with different observation time
  windows $T$.  For each noise spectra, the noise amplitude $I_{1}$
  was chosen so that $F(t=0.1{\textrm ~s})=10$ for all noise spectra.
  This ensures that at $t=0.1$~s the neurons absorb the same amount of
  noise power, despite the fact that the noise spectra are quite
  different. Nevertheless, for times $t$ after $0.1$~s, the
  Fano factor differs considerably for different spectra. This occurs
  because neurons ``integrate noise'' over a bandwidth $\approx 1/(\pi
  t)$ as described by the filter function
  Eq.~(\ref{filterfunc}).\label{fig:fanopower}}
\end{figure}

\subsection{Fano factor in the perfect integrate and fire model: Numerical results}

\begin{figure}
\includegraphics[width = 0.49\textwidth]{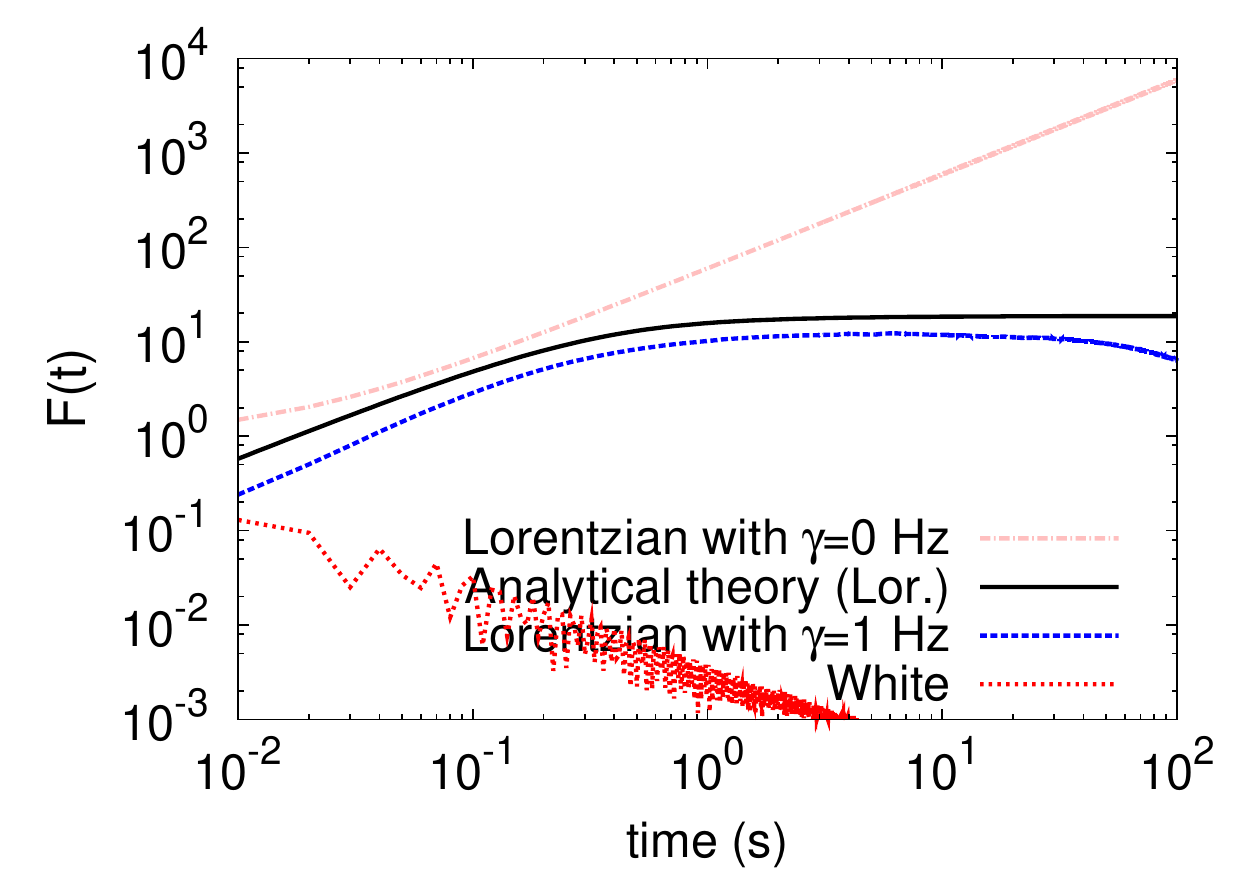}
\caption{(color online) Fano factor [Eq.~(\ref{fano})] for the
  $R=\infty$ (non-leaky or perfect) integrate and fire model as a
  function of time for white noise, Lorentzian noise with
  $\gamma=1$~Hz, and $\gamma=0$~Hz, the static case.  We assumed
  $I_0=I_1=2\times 10^{-10}$~A and other parameters as in
  Table~\ref{table1}.  Also shown is a comparison between the
  numerical simulation and the analytical expression
  Eq.~(\ref{fanoanaly}) for Lorentzian noise.  For white noise, $F(t)$
  tends to a quite small value at long times; for Lorentzian noise, $F(t)$ plateaus
  at times $t\approx (2\pi\gamma)^{-1}$.\label{fig:fanoperf}}
\end{figure}

Figure~\ref{fig:fanoperf} shows the Fano factor, as a function of
time, for $I_0=I_1=2\times 10^{-10}$~A and $T=10^{2}$~s for white
noise, and for Lorentzian noise with $\gamma=1$~Hz [$\tau_c =
1/(2\pi)$~s] and $\gamma=0$~Hz ($\tau_c=\infty$).  We also plot the
analytical expression Eq.~(\ref{fanoanaly}) for the $\gamma=1$~Hz
Lorentzian Fano factor; as expected, the analytical expression shows
slightly higher disorder [larger $F(t)$] than our Lorentzian noise
simulation, because the latter only includes positive input currents
(they are in close agreement when $I_1\ll I_0$). For white noise, the
Fano factor tends to a small value at long times, in accordance with
Eq.~(\ref{finf}) that gives $F(t\rightarrow \infty)\propto
1/\gamma_{\rm{max}}$ ($\gamma_{\rm{max}}=10^{5}$~Hz is the upper
frequency cut-off of our white noise spectrum).  For the $\gamma=0$
Lorentzian, we have $F(t)\propto t$ in accordance with the limit
$\tau_c\rightarrow \infty$ of Eq.~(\ref{fanoanaly}).
 
\begin{figure}
\includegraphics[width = 0.49\textwidth]{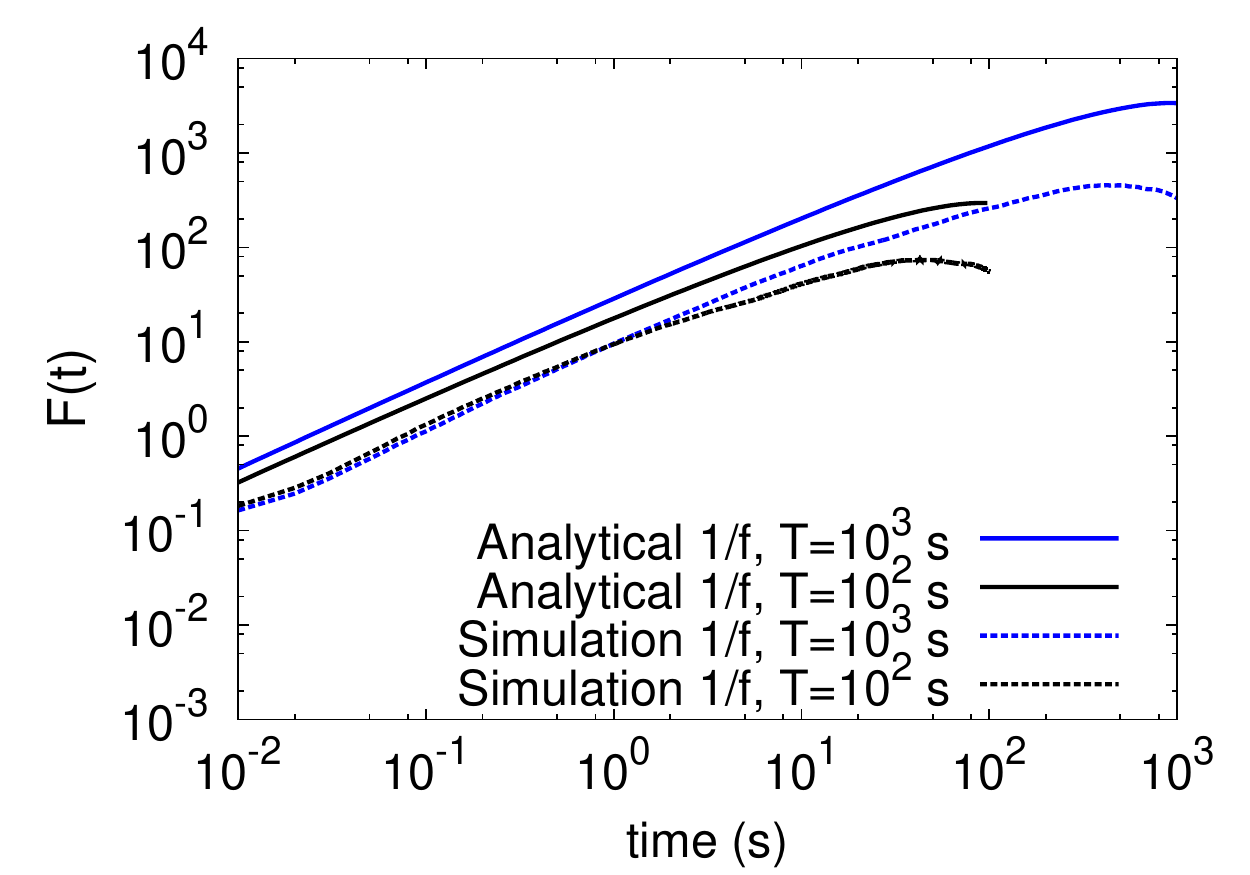}

\caption{(color online) Fano factor [Eq.~(\ref{fano})] for the
  non-leaky (perfect) integrate and fire model as a function of time
  for $1/f$ noise, using the same parameters as in
  Fig.~\ref{fig:fanoperf}. We show simulations of $1/f$ noise for two
  different time series: $T=10^2$~s and $T=10^{3}$~s, where $T$ is the
  length of the simulation time window. We also show the corresponding
  analytical approximations Eq.~(\ref{fanooneoverf}). Note how the
  Fano factor due to $1/f$ noise increases logarithmically until it
  reaches a slight saturation at $t=T$, the ``maximal experimental
  time''. This behavior agrees qualitatively to what is observed in
  measurements on single visual neurons in cats and insects
  \cite{teich97}.\label{fig:fanoperfoneoverf}}
\end{figure}

Figure~\ref{fig:fanoperfoneoverf} shows the neuron Fano factor subject
to $1/f$ noise, using two different simulation time windows:
$T=10^{2}$~s and $T=10^{3}$~s. We also show the corresponding analytic
results using Eq.~(\ref{fanooneoverf}) with $\gamma_{\rm{min}}=T^{-1}$
and $\gamma_{\rm{max}}=N/T$ ($N$ is the number of frequency intervals
used in our simulation, see Appendix~\ref{appendixa}). Similar to the
Lorentzian case, the analytic expressions for $F(t)$ are larger than
the numerical results because the former does not take into account
the step function in Eq.~(\ref{currit}).
  
As expected, we find that the qualitative behavior of 1/$f$ noise is markedly
different from Lorentzian noise. While for Lorentzian noise $F(t)$
increases linearly with $t$ until it reaches an asymptotic maximum at
$t\approx \tau_c$, for $1/f$ noise $F(t)$ increases logarithmically
[$\propto -t\ln{(t/T)}$] and only reaches a slight saturation when
$t\approx T$, the maximum possible value of time.  

\subsection{Fano factor in the leaky integrate and fire model}

\begin{figure*}
  \subfigure[Lorentzian Noise, $\gamma=1$~Hz]{\label{fig:fanolor}\includegraphics[width=0.49\textwidth]{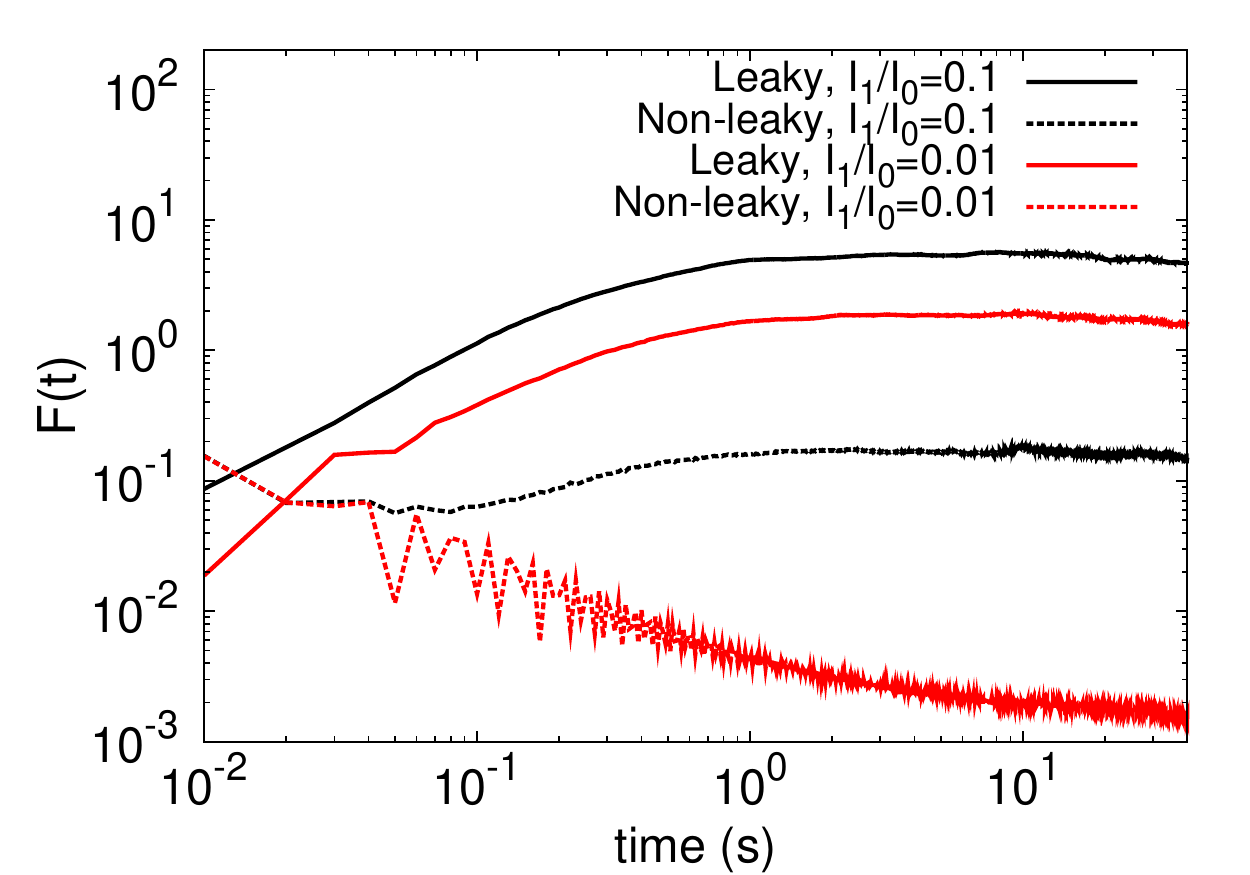}}
  \subfigure[1/$f$ Noise, $T=100$~s]{\label{fig:fanopink}\includegraphics[width=0.49\textwidth]{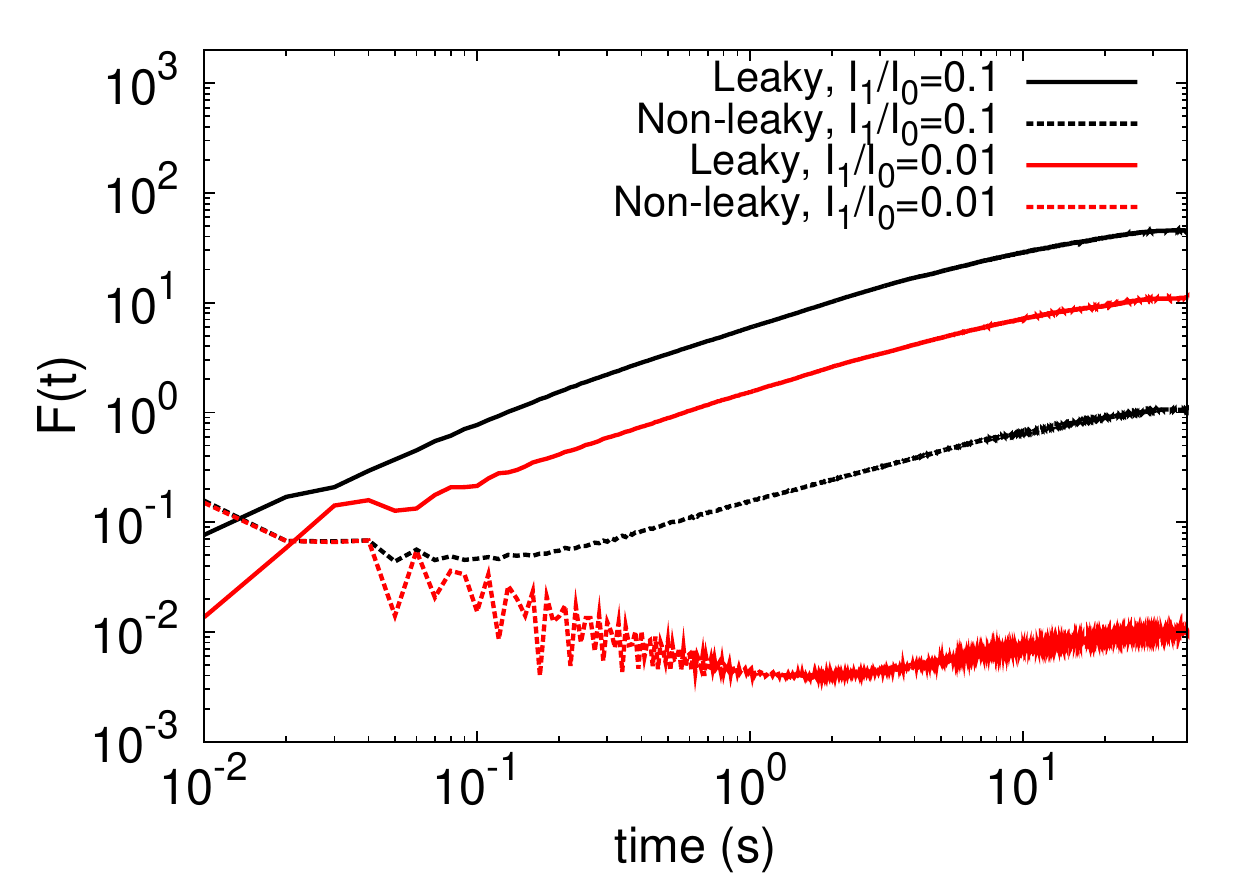}}
  \caption{(color online) Fano factor for integrate and fire neuron
    with leakage and without leakage, for $I_1/I_0= 1, 0.1, 0.01$.
    Leakage increases the Fano factor considerably, and can produce
    large quantitative difference at low levels of noise.
    \label{fig:fanocomp}}
\end{figure*}

We now describe the impact of leakage on the Fano factor.  We
simulated the LIF model under Lorentzian and $1/f$ noise of several
different noise levels. Figure~\ref{fig:fanocomp} compares the Fano factor
with leakage and without leakage; in every case, leakage increases the
Fano factor noticeably (i.e. there is increased variability in the
spike train).  This happens because in the presence of leakage, the neuron
tends to forget past inputs that were not strong enough to break the
threshold barrier and returns to its rest state
even though it received a considerable amount of subthreshold input.
This situation is dramatically different from the perfect (non-leaky)
model: In the absence of leakage, every subthreshold stimulation
increases the charge of the neuron, priming it for firing. The quantitative difference in Fano
factor highlights the importance of leakage in neuron dynamics. For
example, note the dramatic quantitative difference between Fano factor
for leaky and non-leaky cases at the level of 1\% of $1/f$ noise
($I_1/I_0=0.01$).

Our explicit numerical results of Fano factor subject to $1/f$ noise
shows that $F(t)$ \emph{does not saturate at long times}. This
property is consistent with the the direct measurements of neuron Fano
factor presented in Teich {\it et.  al.}  (See Fig.~6 in
\cite{teich97}). Their experimental Fano factor increases well beyond
one, and does not appear to reach a plateau at long times. Note that
previous calculations based on Lorentzian noise
\cite{middleton03,schwalger08} have showed that the Fano factor
saturates at long times.  This result
is independent of whether one uses the perfect or the leaky integrate
and fire model.

One can argue that with respect to a Lorentzian noise model, the
experimental data only explicitly rules out cases with $\tau_{c} < T$,
i.e. Lorentzian models with correlation time longer than the
experimental window will not show saturation. This is true, but the
Lorentzian and $1/f$ model predictions for $F(t)$ differ in other
respects as well, which do not depend on $\tau_c$ or $T$.

The Fano factor $F(t)$ for the experimental data has a tangent slope
of $0.4-0.5$ around $t\sim 1$~s in log-log plot (Fig.~6 of
\cite{teich97}). The predicted slope of the Lorentzian Fano factor,
however, is $1$ for $t\ll \tau_c$ (before saturation takes place).
This difference rules out the possibility that the experimental
results can be understood in terms of an unsaturated Lorentzian model.

The Fano factor under $1/f$ noise has a tangent slope of approximately
$0.7$ around $t\sim 1$~s
(Figs.~\ref{fig:fanoperfoneoverf}~and~\ref{fig:fanocomp}b). Note that
the Fano factor increases logarithmically, so the exponent depends on
particular time $t$ chosen to measure the slope.  We repeated our
simulation using $1/f^{0.6}$ noise, and obtained results very similar
to Figs.~\ref{fig:fanoperfoneoverf}~and~\ref{fig:fanocomp}b, except
that the slope was reduced to $\approx 0.5$. This shows that a LIF
model subject to $1/f^{\alpha}$ noise can provide an excellent fit to
experimental data of long time neuron dynamics, and further suggests
that neuron input noise is better approximated by a $1/f$-like
spectrum, than a Lorentzian with a single characteristic correlation
time.

\section{How neurons respond to a sudden step excitation\label{suddensec}}

\begin{figure}
\includegraphics[width = 0.49\textwidth]{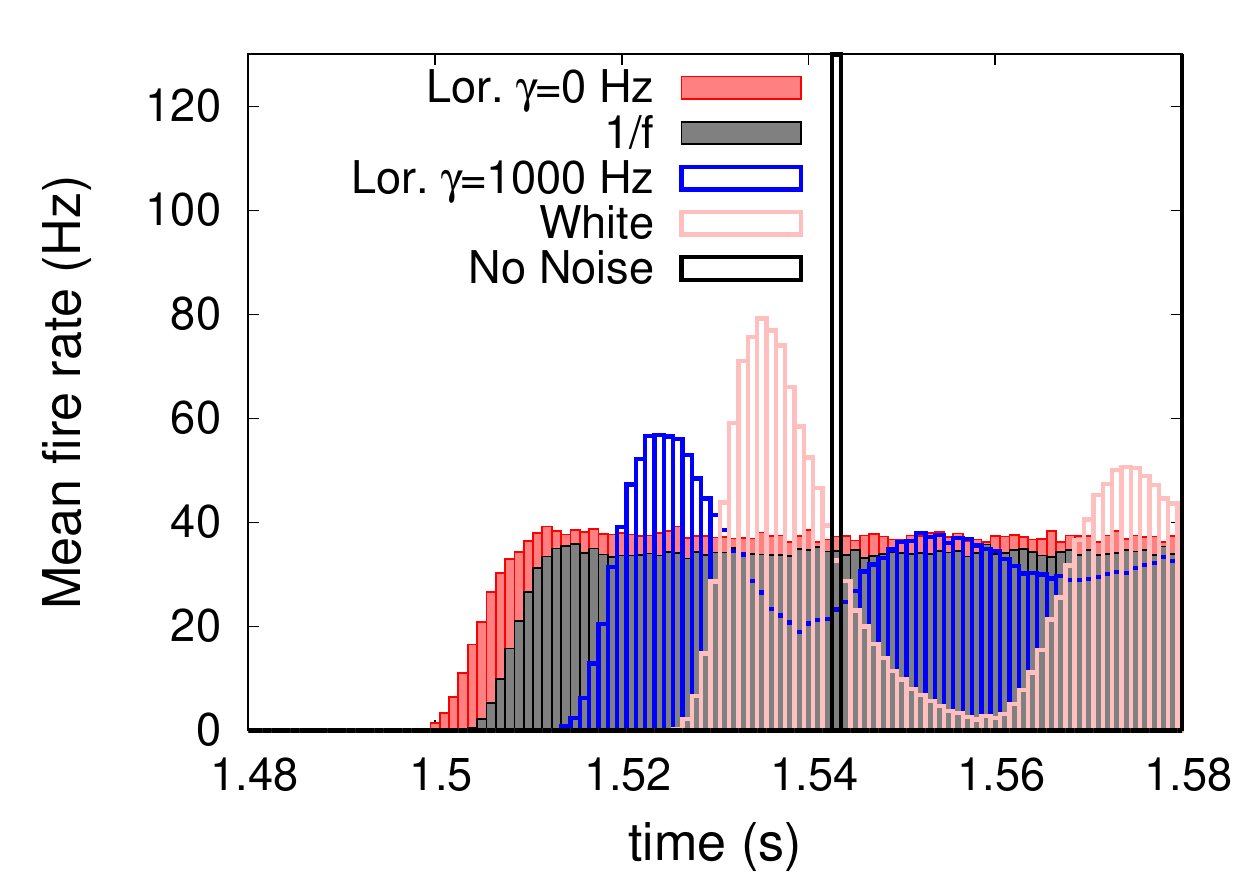}
\caption{(color online) Mean fire rate of single neuron in response to
  step excitation under various types of noise: Lorentzian noise with
  $\gamma=0$~Hz (``static'' case), Lorentzian noise with
  $\gamma=1000$~Hz, white noise, and $1/f$ noise.  We also included
  the noiseless case for comparison.  The presence of noise primes the
  neurons to react significantly faster. The reaction time is optimal
  (almost instantaneous) for the ``static'' case.  Surprisingly, under
  $1/f$ noise the neuron response time is nearly
  optimal.\label{fig:stepmfr}}
\end{figure}

We now analyze the impact of low frequency noise on the reaction time of a LIF neuron. 
We consider the reaction of a neuron subject to the following stimulus,
\begin{equation}
I(t)=\theta\left(I_1 \eta(t) + I_0 \theta(t-t_{\rm{step}})\right).
\label{step}
\end{equation}
For times between $t=0$ and $t=t_{\rm{step}}$, the current input is
pure noise with subthreshold amplitude $I_1$; at $t=t_{\rm{step}}$, a
superthreshold bias current $I_0$ is suddenly turned on in addition to
the noise.  In the simulations below we used $t_{\rm{step}}=1.5$~s,
$I_0=4.3\times 10^{-10}$~A, and $I_1=0.3 I_0$.
Figure~\ref{fig:stepmfr} shows the mean fire rate averaged on $1$~ms
bins after 100,000 time series are taken into account.  The ensemble
of time series can be thought of either an actual ensemble of
different neurons or to a single neuron subject to statistically
similar excitations at different times.  In both cases, we can make
claims of optimality on ``average''.

In the absence of noise, the LIF neuron takes $43$~ms to respond. 
The quickest response is
obtained in the case of Lorentzian noise with $\gamma=0$~Hz, that is
equivalent to $\tilde{S}(f)=\delta(f)/(2\pi)$, the static limit
discussed in section~\ref{intronoise}. Note that in this case the
noisy current does not change in time, and is equivalent to a bias
current with amplitude picked from a Gaussian distribution.
\emph{The response under $1/f$ noise lags behind the
  static case by only $\approx 5$~ms, i.e., it is nearly optimal}.
Both static and $1/f$ noise reach their steady state much faster when
compared to other types of noise. Lorentzian noise with roll-off
frequency $\gamma=1000$~Hz is intermediate between white and $1/f$
noise. Clearly, the response time improves as the noise gets dominated
by low frequency components.

\begin{figure}
\includegraphics[width = 0.49\textwidth]{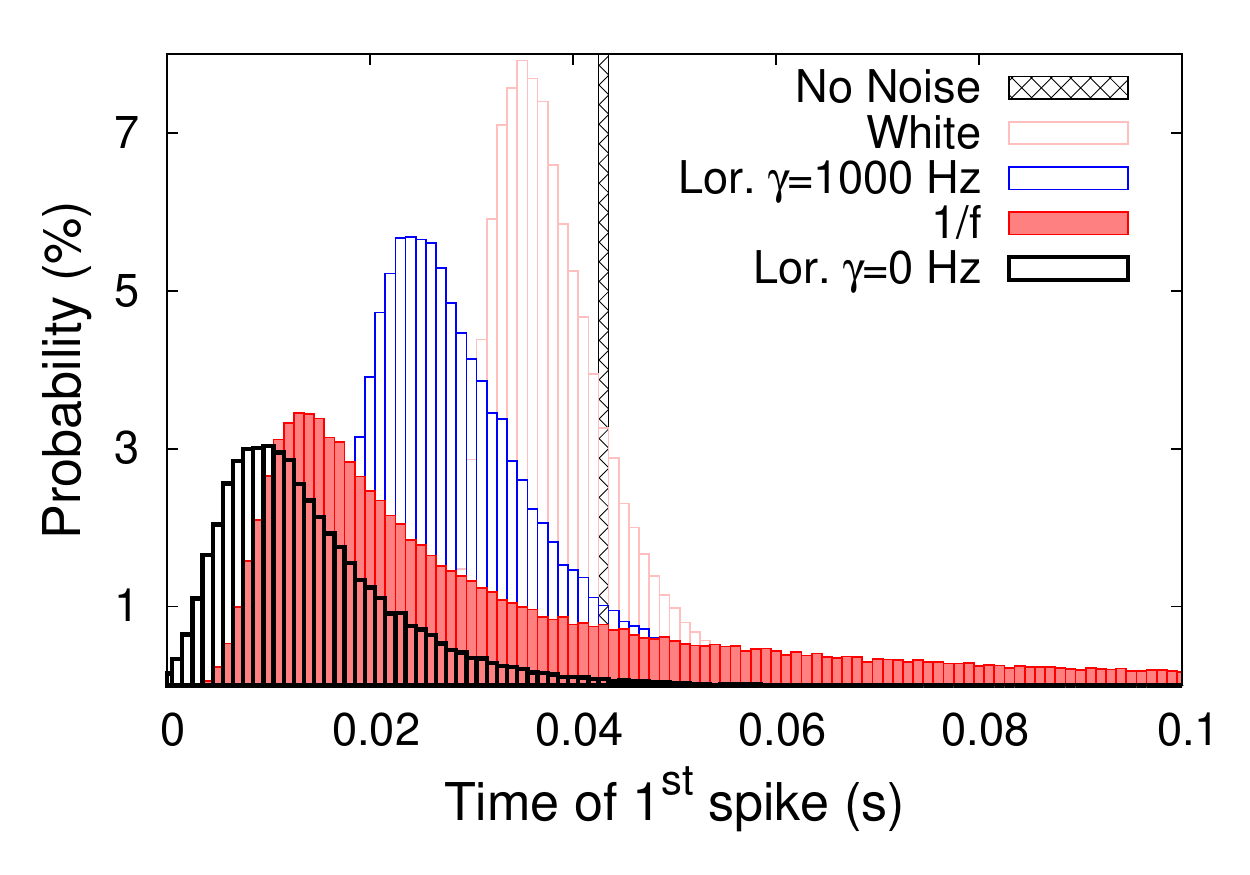}
\caption{(color online) Histogram of first spike times under a step excitation
  subject to various types of noise. In the case of white noise, it
  takes $30$~ms for 1\% of the neurons to spike for the first time.
  The situation is markedly different for the static case (Lorentzian
  with $\gamma=0$~Hz) and for $1/f$ noise. Here some neurons respond
  almost immediately, while others take a long time to spike; note the
  long time tail in the distributions for Lor. 0~Hz and $1/f$ noise.
\label{fig:firststep}}
\end{figure}

In order to elucidate the mechanism by which noise sensitizes neuron
response time, we present two additional figures.
Figure~\ref{fig:firststep} shows a histogram of first spike times.
Under white noise conditions, it takes approximately $30$~ms for 1\% of the
neurons to spike for the first time.  On the other hand, it takes
only $8$~ms for 1\% of the $1/f$ noise neurons to spike and a mere $2.5$~ms for 1\% of static noise
(i.e.~Lorentzian $\gamma=0$) neurons to spike for the first time.
Neurons subject to $1/f$ noise and the $\gamma=0$ static noise have a
much wider distribution of first spike times.  Some neurons spike
almost immediately, while others take a long time to spike -- note the
extended tail in the distributions. 
This broad distribution implies a greater variability 
in the neuronal interspike interval and a degradation of any information coded therein. 
Extreme low frequency noise ($1/f$ and $\gamma=0$) result in a trade off between reliability and
rapid response.

\begin{figure}
\includegraphics[width = 0.49\textwidth]{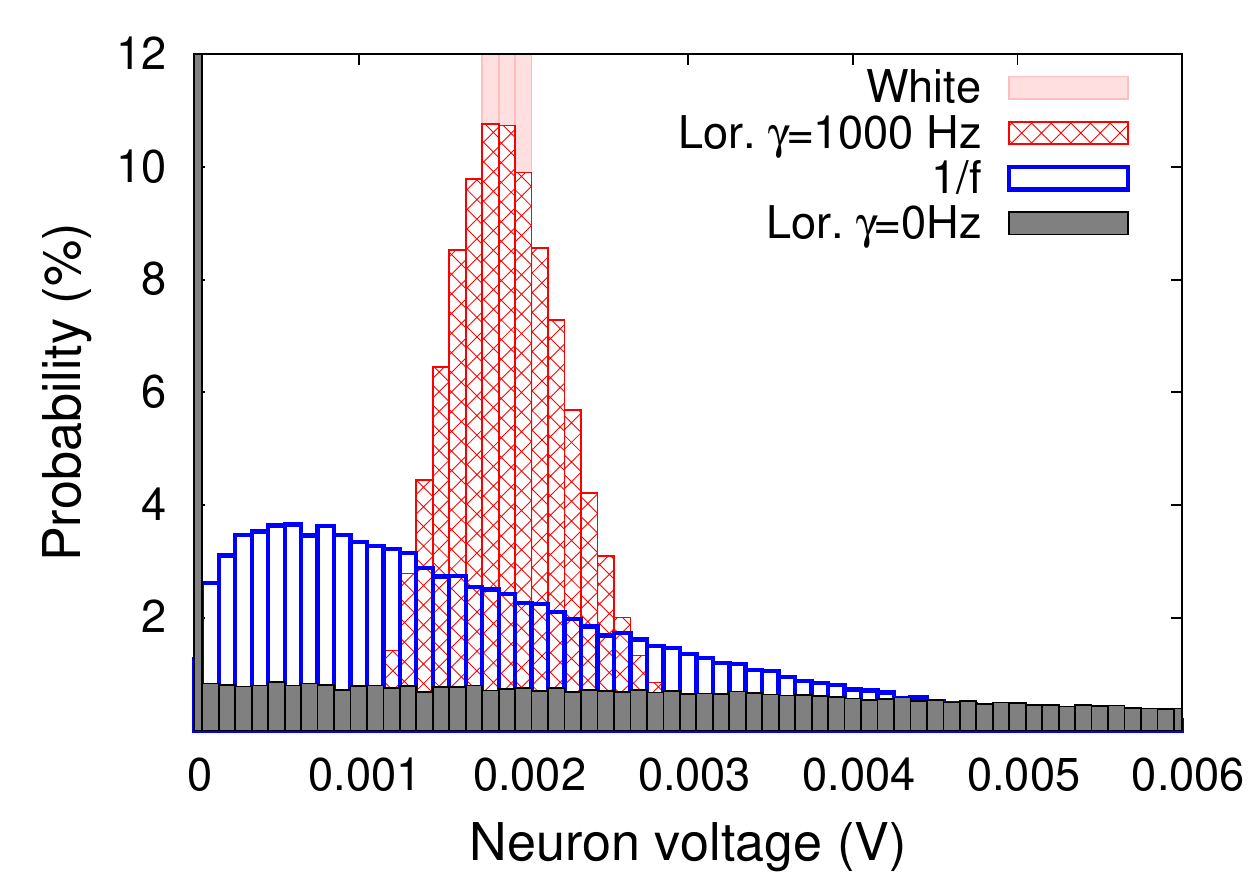}
\caption{(color online) Neuron voltage right before the step
  excitation arrives. Note that for the static case (Lor. 0~Hz) and
  for the $1/f$ noise case, a significant fraction of neurons have
  quite high voltage. Hence these neurons will reach threshold much
  faster, explaining how noise can ``prime'' neurons for a fast
  response time.\label{fig:prestepvoltage}}
\end{figure}

Figure~\ref{fig:prestepvoltage} sheds light on the origin of this
effect, by plotting the distribution of neuron voltages just before
the step stimulus is applied. Here we see why the static case is optimal:
The distribution of neuron voltages is nearly flat, and extends close
to threshold. Hence, when the stimulus is applied, a significant
amount of ``primed neurons'' will reach threshold almost instantly.
While the voltage distribution for $1/f$ noise is not flat, it is
broad and extends all the way to high voltages. Similar to the static
case, the presence of a tail extending near threshold implies that a
significant number of neurons are ``primed'' by $1/f$ noise; these
neurons will react nearly instantly to the stimulus. 

We note in passing that it is possible to engineer a Lorentzian noise
spectrum to yield a response time similar to that for $1/f$ noise via
an appropriate choice of $\gamma=1/T$, where we recall that $T$ is the
simulation time window. For $T=2$~s, we find that $\gamma=0.5$~Hz does
the trick, in agreement with expectations based on our previous
analysis of the distribution of interspike time intervals for
different conditions (c.f. Fig.~\ref{fig:ISI}): The shortest
interspike time interval for a Lorentzian with $\gamma=0.5$~Hz is
similar to that for $1/f$ noise.  The two distributions are not
identical. The ISI for the $1/f$ has a tail that extends to much
longer times.


\section{Conclusions\label{concsec}}

In conclusion, we presented analytical and numerical calculations of
the perfect and the leaky integrate and fire neuron aimed at
elucidating the impact of $1/f$ noise on single neuron dynamics.
Though more difficult to analyze than the perfect integrate and fire
model that is commonly used in noise and network studies, our LIF
model is a more realistic model of a neuron and as we have shown, the
inclusion of the ``leakiness'' gives rise to much higher disorder.

With regard to the response of the LIF to $1/f$ noise, we find a 
surprising dichotomy: While it degrades the ability to transmit
information using interspike times, it manages to enhance the overall 
response time (of an ensemble of neurons) to a sudden stimulus by ensuring
that a subset of neurons are primed with a near threshold voltage.

Our explicit numerical simulations of neuronal response to a sudden
step excitation elucidates the mechanism by which noise can enhance neuron response time.  
Neuron response times were shown to be
optimal under static noise (noise sharply peaked at zero frequency)
because a large number of neurons are close to threshold just before
the step stimulus arrives. This select group of neurons act as the
alarm bells: They respond almost instantly to the step stimulus.
Remarkably, as we demonstrated, the case of $1/f$ noise is not much
different; the distribution for neuron voltages just before stimulus
also has a long tail extending towards threshold.  These results
allude to a possible explanation for why the brain is populated by an
astronomically large number of neurons. It may well be an evolutionary
adaptation designed to take advantage of the ambient noise to enhance
the probability of survival.  Neuron redundancy enables faster
response times in the presence of low frequency noise, which in turn
allows an animal to react quickly to a sudden danger.

However, the apparently beneficial feature noted above does not come
without cost. The $1/f$ noise trades off speed for reliability by
introducing much more variability in the properties of the resulting
spike train. We quantify this uncertainty using the Fano factor.  Our
analysis of the Fano factor reveals that in the presence of $1/f$
noise, this measure of disorder increases logarithmically as a
function of time.  On a positive note, we find an excellent
qualitative agreement between Fano factor for the $1/f$ noise and the
Fano factor derived from laboratory results of experiments with single
neurons \cite{teich97,turcott95}.  Specifically, the latter also rises
monotonically well beyond one and shows no evidence for saturation.
This agreement suggests that the neuron input noise is better
approximated by a scale-free $1/f$-like spectrum than the more
commonly invoked low frequency Lorentzian spectrum.

The rate at which the Fano factor for a Lorentzian spectrum grows with
time prior to saturation is $F(t)\propto t$, independent of $\gamma$.
Moreover, the Lorentzian Fano factor always tends to a plateau at long
times \cite{middleton03, schwalger08}.  Both tendencies are at odds
with the behavior of the experimentally measured Fano factor
regardless of whether the Lorentzian spectrum is fine-tuned to yield a
response time similar to that of $1/f$ noise.  We note that these
claims can be definitively tested by repeating the experiments of
Teich {\it et al.}\cite{teich97} with increasingly longer experimental
time windows $T$.

The logarithmically rising Fano factor reflects the fact that the long time spike dynamics is dominated either by
periods of extended inactivity or by periods of aggressive bursting.
This behavior is due to the lack of ergodicity in $1/f$ noise, i.e.,
the fact that it lacks a characteristic correlation time.  Not
surprisingly, therefore, the neuron dynamics in the presence of $1/f$
noise is very different from that due to Lorentzian noise.  As an
aside, we note that the degree of uncertainty is also substantially
greater in the leaky model than in the non-leaky (perfect) case.

These conclusions are consistent with the observation that some
neurons seem to spike in a very irregular fashion.  The temporal gaps
of a spike train has much larger information capacity and for this
reason, there is considerable body of work arguing that neurons use
the timing intervals to encode information.  Loss of reliability due
to low frequency noise, however, limits the information capacity of
the spike trains \cite{rossum03}.  On the other hand, research has
shown that in certain cases neurons can spike with high degree of
reproducibility \cite{steveninck97}.  Whether the origin of highly
reproducible spike patterns is due to extremely low noise at the
single neuron level, or due to a network effect that compensates for
the noise remains to be seen.  A more interesting possibility is that
the various functional regions of the brain may have evolved different
strategies for managing ambient noise, depending on function and
associated information capacity demands.

\begin{acknowledgments}
We wish to thank the Natural Sciences and Engineering Research Council of Canada for supporting this work. 
\end{acknowledgments}

\appendix

\section{Numerical simulation of Gaussian noise with arbitrary spectral density\label{appendixa}}

To simulate the noise used in Eq.~(\ref{currit}) we used a variation
of the efficient algorithm proposed by Timmers and Koenig
\cite{timmer95}.  Consider the time window from $t=0$ to $t=T$. Define
a discrete set of $N$ time instants $t_m=n \Delta t/2$, where
$n=0,1,\ldots,2 N-1$ and $\Delta t=T/N$. Choosing $N$ as a power of
$2$ allows the use of the fast Fourier transform algorithm, with
significant speed up.  The associated set of ``lower half''
frequencies are $f_m=m/T$, with $m=0,1,\ldots, N$, and the ``upper
half frequencies'' are $f_m=(2N-m)/T$ for $m=N+1,N+2,\ldots, 2N-1$. We
are now ready to state the algorithm that generates individual real-valued time
series $\eta(t_m)$ ($\tilde{\eta}(f)$ are their Fourier transforms):

\begin{enumerate}

\item Set $\tilde{\eta}(f_0)=0$;

\item For each $m=1,\ldots, N-1$, set $\tilde{\eta}=\textrm{e}^{i\pi
    r_m} \sqrt{\tilde{S}(f_m)}$, where $r_m$ is a random number in the
  interval $[0,1)$;

\item Set $\tilde{\eta}(f_N)=\sqrt{\tilde{S}(f_N)}$;

\item Set $\tilde{\eta}(f_{N+m})$ equal to the complex conjugate of $\tilde{\eta}(f_{N-m})$ for all $m=1,\ldots, N$;

\item Finally, take the inverse Fourier transform of
  $\tilde{\eta}(f_m)$. The resulting $\eta(t_m)$ realizes an
  individual time series of the Gaussian process with noise spectrum
  $\tilde{S}(f_m)$.

\end{enumerate}

\begin{figure*}
  \subfigure[White Noise]{\label{fig:wtime3}\includegraphics[width=0.32\textwidth]{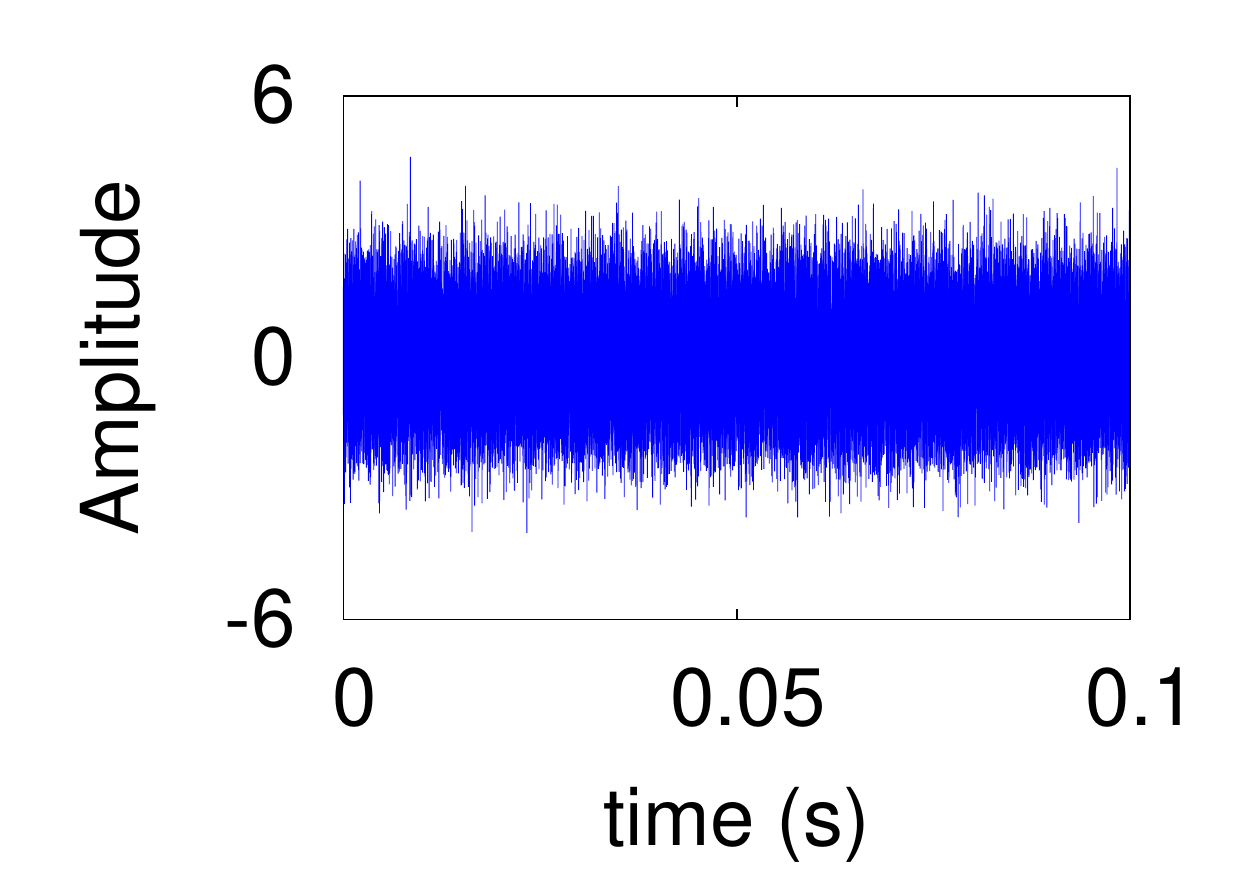}}
  \subfigure[1/$f$ Noise]{\label{fig:ptime3}\includegraphics[width=0.32\textwidth]{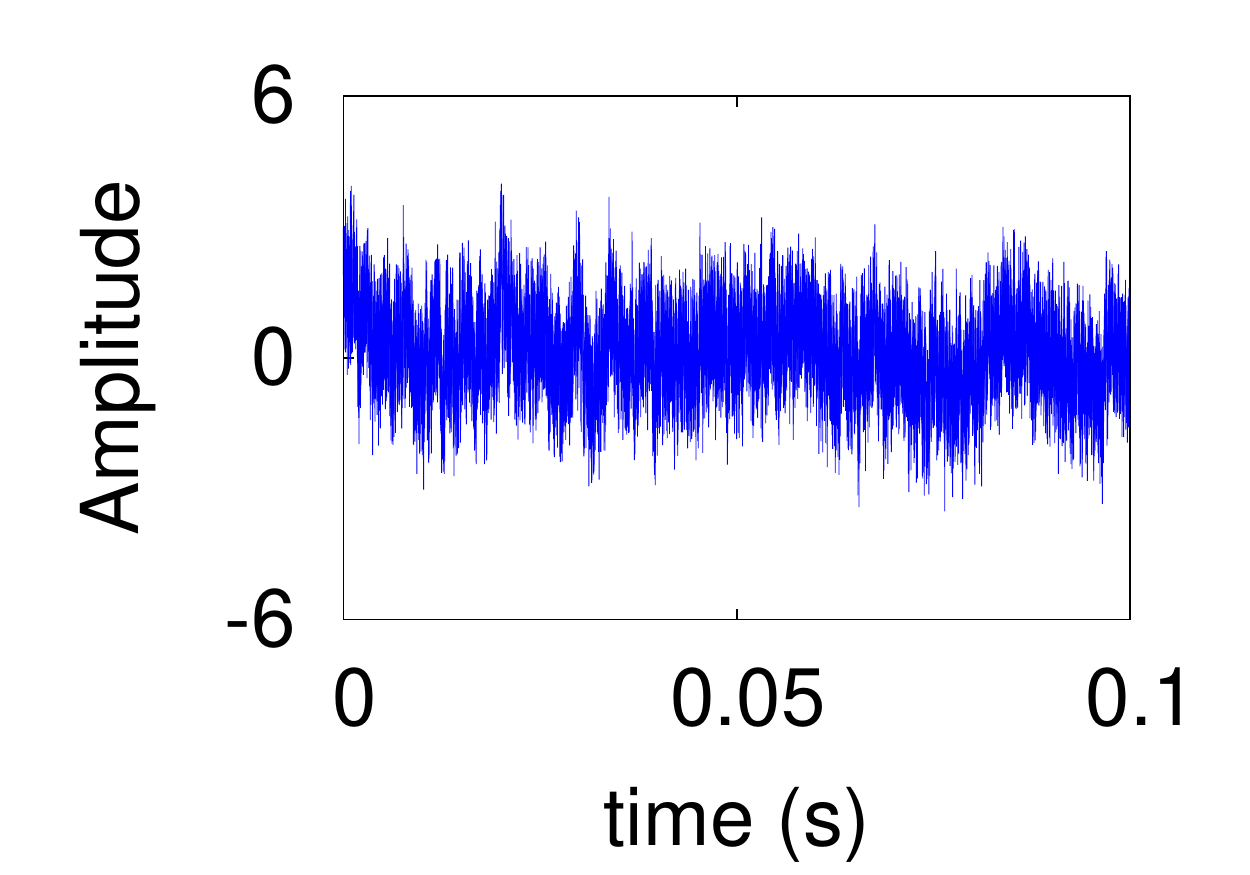}}
  \subfigure[Lorentzian Noise, $\gamma=10$~Hz]{\label{fig:ltime3}\includegraphics[width=0.32\textwidth]{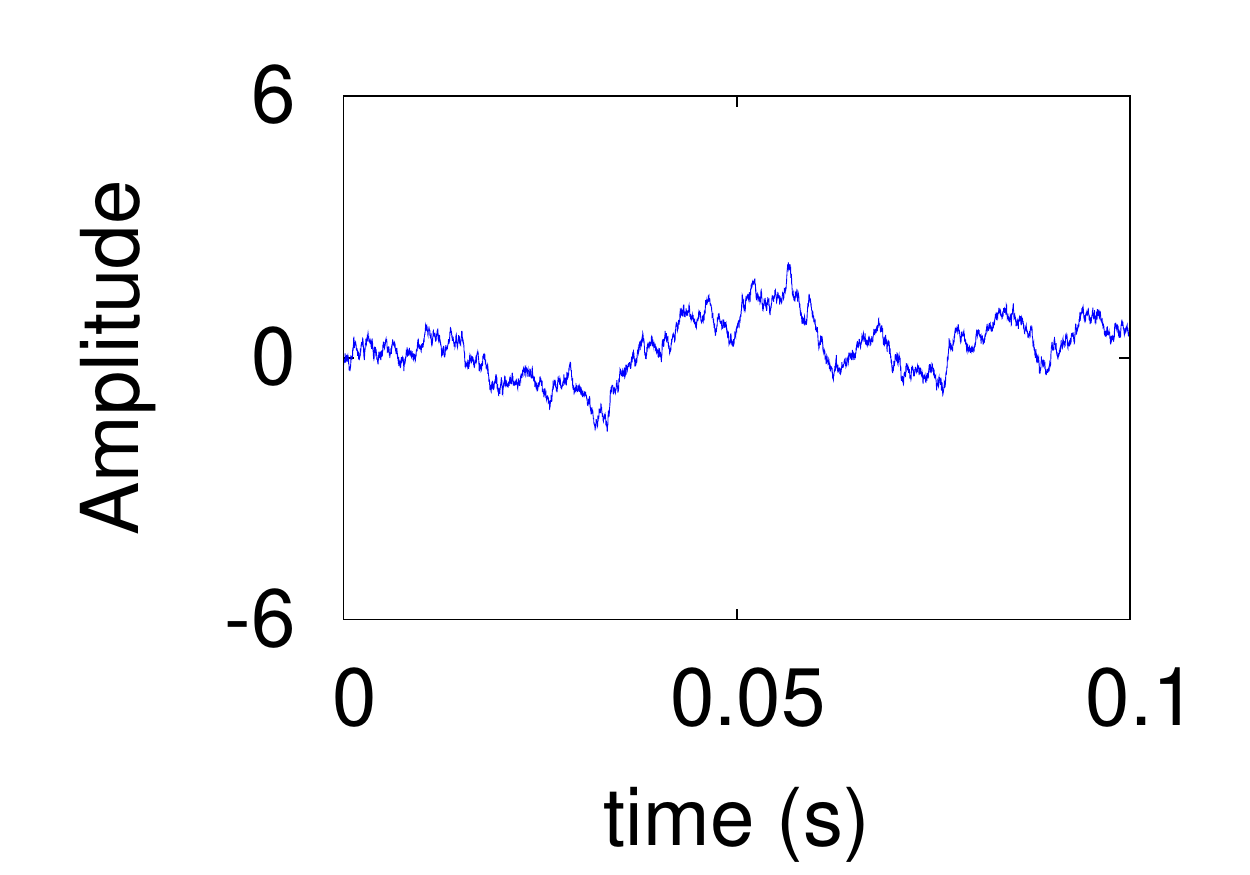}}
  \caption{(color online) Example time Series for three different kinds of noise spectra.\label{fig:timeseries3}}
\end{figure*}

\begin{figure*}
  \subfigure[White Noise]{\label{fig:wdist}\includegraphics[width=0.32\textwidth]{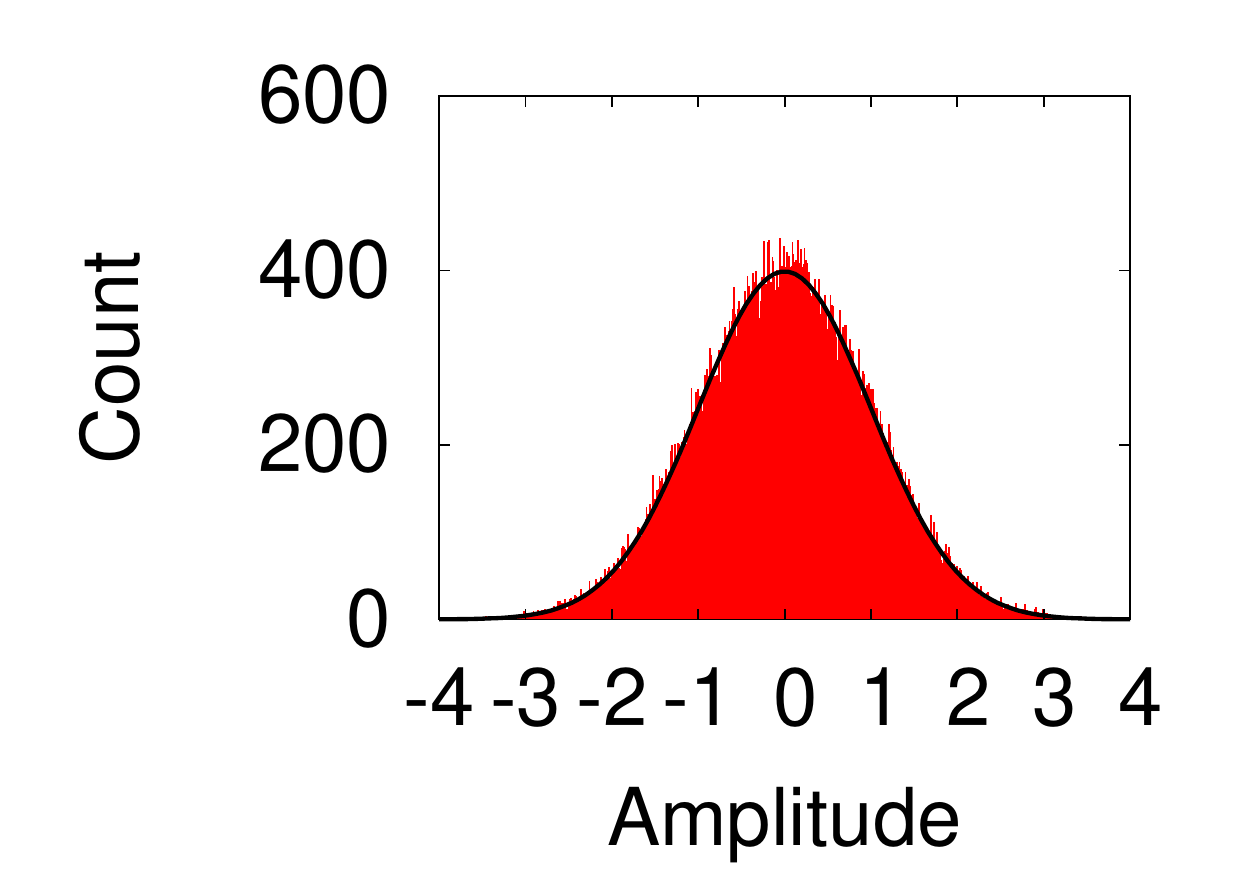}}
  \subfigure[1/$f$ Noise]{\label{fig:pdist}\includegraphics[width=0.32\textwidth]{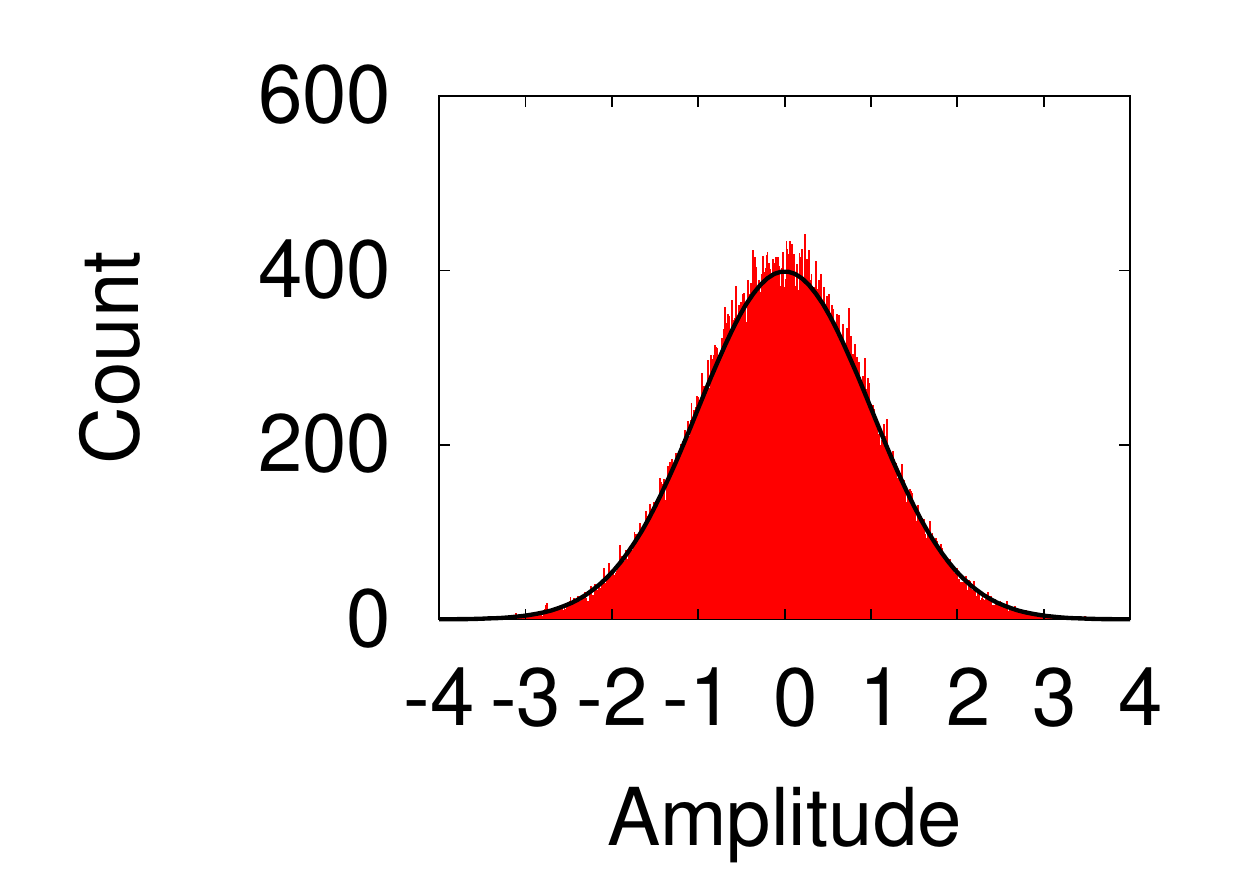}}
  \subfigure[Lorentzian Noise, $\gamma=10$~Hz]{\label{fig:ldist}\includegraphics[width=0.32\textwidth]{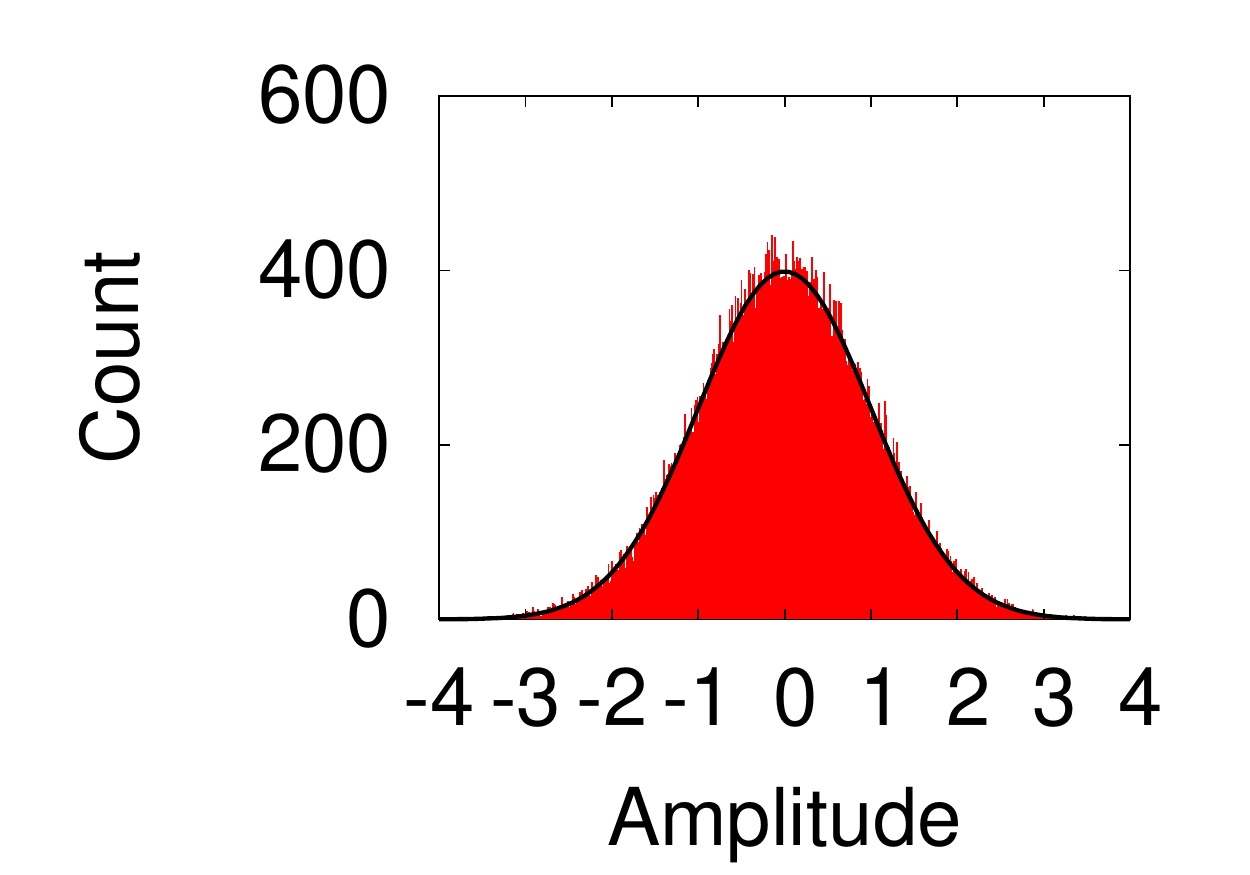}}
  \caption{(color online) Histogram for the amplitudes of $\eta(t)$ for 100,000 time
    series, demonstrating that $\eta(t)$ is Gaussian distributed.\label{fig:ampldist}}
\end{figure*}

\begin{figure*}
  \subfigure[White Noise]{\label{fig:wcorr}\includegraphics[width=0.32\textwidth]{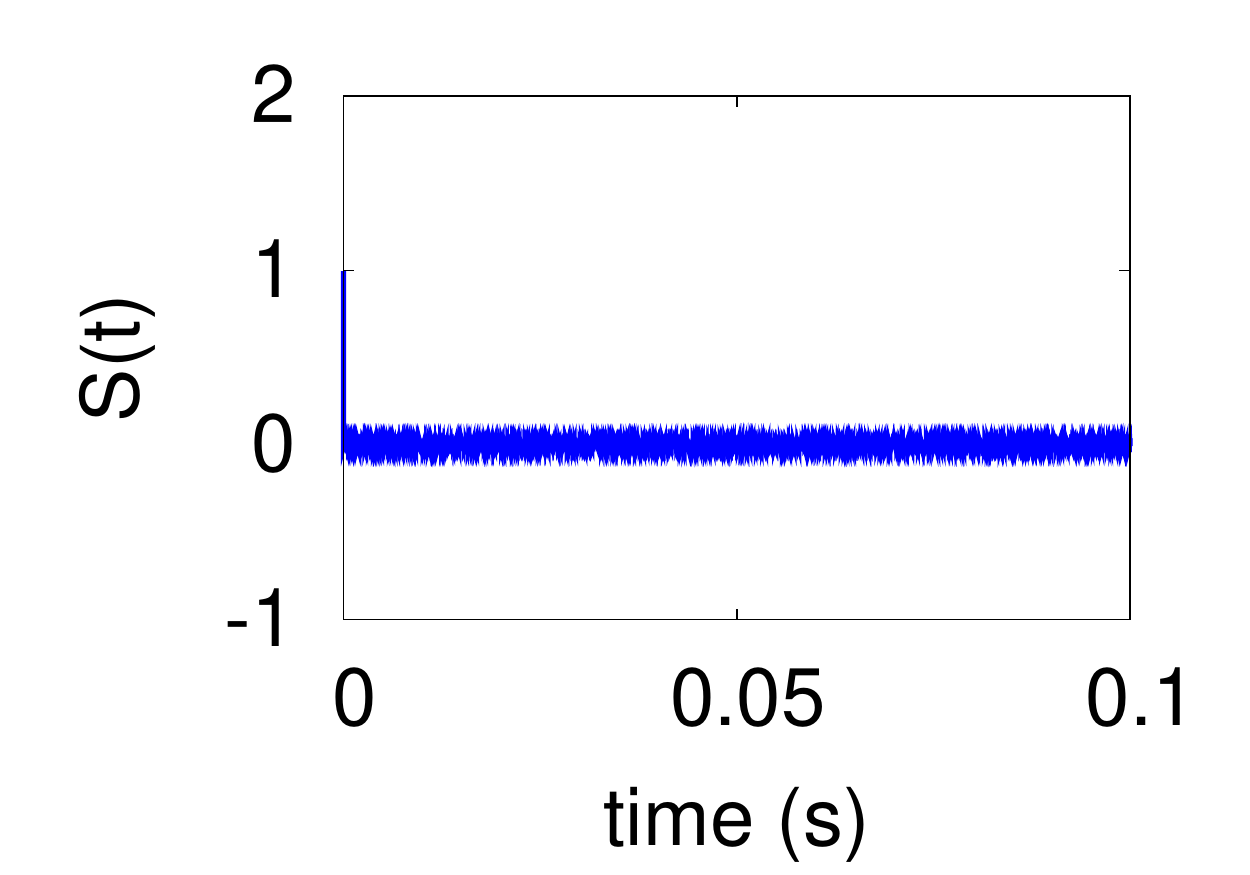}}
  \subfigure[1/$f$ Noise]{\label{fig:pcorr}\includegraphics[width=0.32\textwidth]{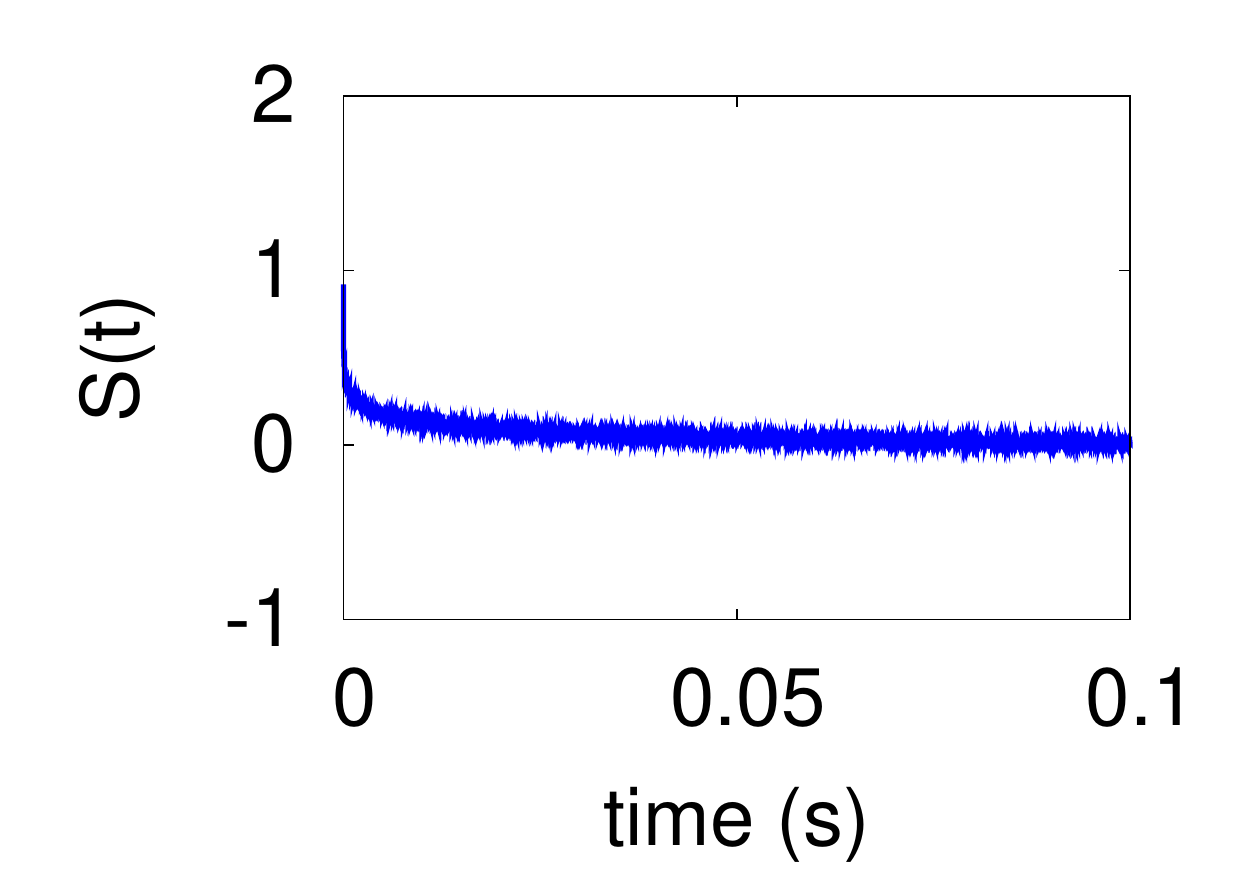}}
  \subfigure[Lorentzian Noise, $\gamma=10$~Hz]{\label{fig:lcorr}\includegraphics[width=0.32\textwidth]{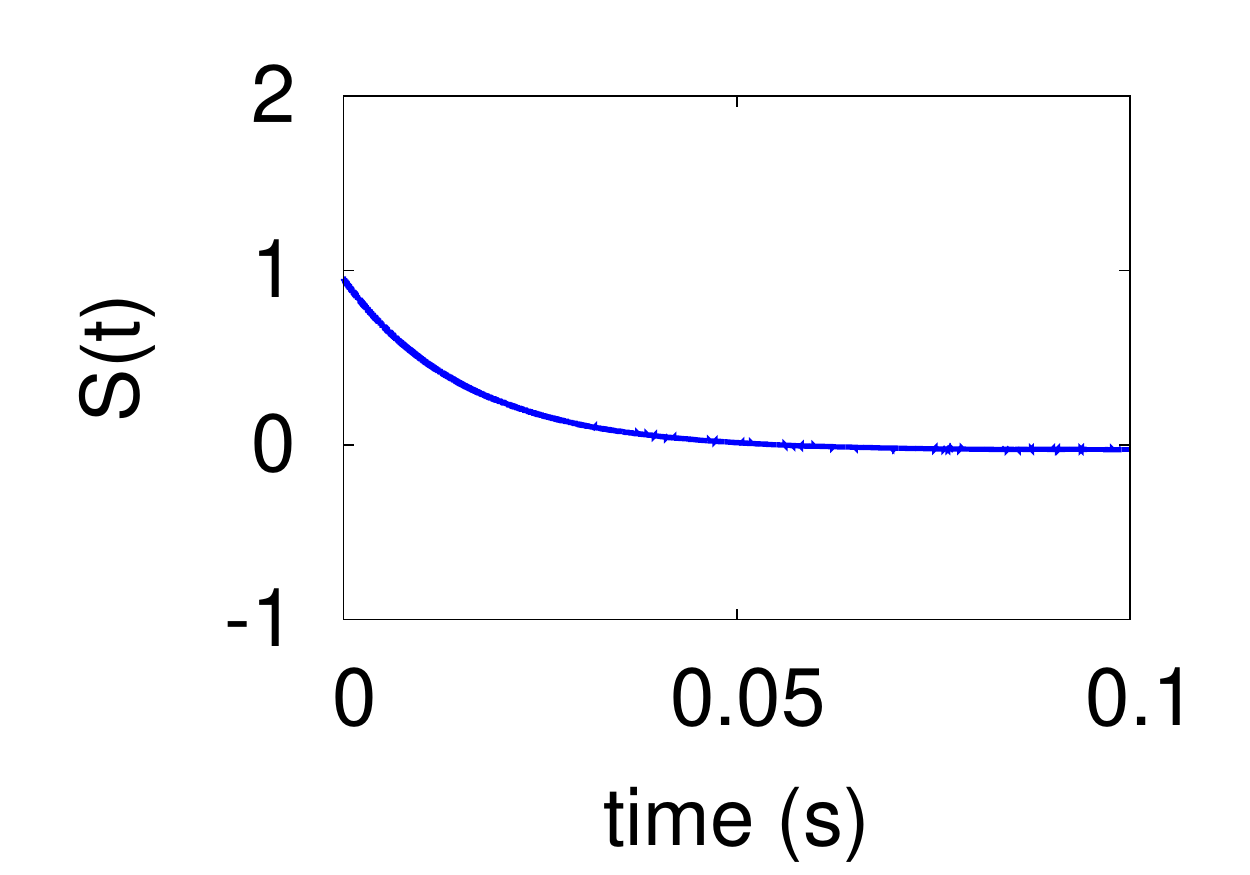}}
  \caption{(color online) Ensemble average correlation function, $S(t)=\langle
    \eta(t)\eta(0)\rangle$, calculated as an arithmetic average of
    several time series generated by the method.\label{fig:corr}}
\end{figure*}

Figure~\ref{fig:timeseries3} depicts three example time series: white
noise, $1/f$ noise, and Lorentzian with half-width $\gamma=10$~Hz. We
simulated 100,000 of these time series and studied their amplitude
distribution and noise spectra [Eq.~(\ref{noisespec})].
Figure~\ref{fig:ampldist} demonstrates that the noise amplitudes are
distributed according to a Gaussian, and Fig.~\ref{fig:corr} computes
the ensemble average of their correlation function, $S(t)=\langle
\eta(t)\eta(0)\rangle$. The latter have the expected forms: For white
noise, $S(t)=\sin{(2\pi
  \gamma_{\rm{max}}t)}/(2\pi\gamma_{\rm{max}}t)$, for $1/f$ noise,
$S(t)\approx 1- [C_E
+\rm{ln}(\gamma_{\rm{max}}t)]/\rm{ln(\gamma_{\rm{max}}/\gamma_{\rm{min}})}$
($C_E=0.5772$ is the Euler-Mascheroni constant), and
$S(t)=\textrm{e}^{-\gamma t}$ for Lorentzian noise.

\section{Exact calculation of the interspike interval histogram for the LIF model subject to static noise\label{appendixb}}

The case of static noise $\tilde{S}(f)=\delta(f)/2\pi$ (a Lorentzian
with $\gamma\rightarrow 0$) is particularly simple because the
stochastic process $\eta(t)$ randomizing the current
Eq.~(\ref{currit}) does not change in time. Each 
$\eta$ is picked from
a Gaussian distribution at $t=0$,
\begin{equation}
p(\eta)=\frac{1}{\sqrt{2\pi}}\textrm{e}^{-\frac{1}{2}\eta^{2}}.
\end{equation}
As a result, the quasi-static method
for calculating the ISI distribution developed in
Refs.~\cite{middleton03}~and~\cite{lindner04} becomes exact. As we
show below, this allows us to compute the ISI distribution exactly even
in the presence of leakage ($R<\infty$).

In the presence of leakage, the voltage is obtained by solving Eq.~(\ref{dvdt}) under a constant current $I=I_0+I_1\eta$,
\begin{equation}
V(t)=R(I_0 +I_1\eta)\left(1-\textrm{e}^{-\frac{t}{RC}}\right).
\end{equation}
The interspike time interval $l$ will be given by the time it takes for this voltage to reach $V_{\rm{th}}$, leading to
\begin{equation}
l=\tau_r - RC \ln{\left(
1-\frac{V_{\rm{th}}/R}{I_0+I_1\eta}
\right)},
\end{equation}
where $\tau_r$ is the refractory time period. 
The ISI distribution can now be computed from the expression
\begin{equation}
P(l)=\frac{1}{N_\eta} \int_{-\infty}^{\infty}d\eta p(\eta) \delta\left[
l-\left(\tau_r - RC \ln{\left(
1-\frac{V_{\rm{th}}/R}{I_0+I_1\eta}
\right)}\right)\right], 
\end{equation}
where the normalization factor 
\begin{equation}
N_\eta= \int_{\left[V_{\rm{th}}/(RI_1)-I_0/I_1\right]}^{\infty} d\eta p(\eta)
\end{equation}
ensuring that $\eta$ is strong enough to ``click'' the delta function. 
After some algebra we obtain the following exact result:
\begin{eqnarray}
P(l)=\frac{CV_{\rm{th}}}{\sqrt{2\pi}I_1 N_\eta}\frac{\textrm{e}^{-\frac{(l-\tau_r)}{RC}}
\textrm{e}^{-\frac{1}{2}\left(\frac{I_0}{I_1}\right)^2
\left[1-\frac{CV_{\rm{th}}/I_0}{RC \left(1-\textrm{e}^{-\frac{(l-\tau_r)}{RC}}\right)}\right]^2}
}{\left[RC \left(1-\textrm{e}^{-\frac{(l-\tau_r)}{RC}}\right)\right]^2}.
\label{isiexact}
\end{eqnarray}
This expression is plotted in Fig.~\ref{fig:ISI}.

\end{document}